%% file: bjsutlieff_HD1160_astro.tex
\documentclass[fleqn,usenatbib]{mnras}

\usepackage{newtxtext,newtxmath}
\usepackage{CJK}
\usepackage{ae,aecompl}

\DeclareRobustCommand{\VAN}[3]{#2}
\let\VANthebibliography\thebibliography
\def\thebibliography{\DeclareRobustCommand{\VAN}[3]{##3}\VANthebibliography}

\usepackage{graphicx}	
\usepackage{amsmath}	
\usepackage{orcidlink}

\title[Characterization of the HD 1160 system]{Exploring the directly imaged HD~1160 system through spectroscopic characterization and high-cadence variability monitoring} 

\author[B. J. Sutlieff et al.]{Ben J. Sutlieff,$^{1,2,3,4}$\thanks{E-mail: ben.sutlieff@roe.ac.uk}
Jayne L. Birkby,$^{5}$
Jordan M. Stone,$^{6}$
Annelotte Derkink,$^{3}$
Frank Backs,$^{7,3}$
\newauthor David S. Doelman,$^{4,8}$
Matthew A. Kenworthy,$^{4}$
Alexander J. Bohn,$^{4}$
Steve Ertel,$^{9,10}$
Frans Snik,$^{4}$
\newauthor Charles E. Woodward,$^{11}$
Ilya Ilyin,$^{12}$
Andrew J. Skemer,$^{13}$
Jarron M. Leisenring,$^{9}$
Klaus G. Strassmeier,$^{12}$
\newauthor Ji Wang (王吉),$^{14}$
David Charbonneau,$^{15}$
and Beth A. Biller$^{1,2}$
\\
$^{1}$Scottish Universities Physics Alliance, Institute for Astronomy, University of Edinburgh, Royal Observatory, Blackford Hill, Edinburgh, EH9 3HJ, UK\\
$^{2}$Centre for Exoplanet Science, University of Edinburgh, Edinburgh, EH9 3HJ, UK\\
$^{3}$Anton Pannekoek Institute for Astronomy, University of Amsterdam, Science Park 904, 1098 XH Amsterdam, The Netherlands\\
$^{4}$Leiden Observatory, Leiden University, P.O. Box 9513, 2300 RA Leiden, The Netherlands\\
$^{5}$Astrophysics, University of Oxford, Denys Wilkinson Building, Keble Road, Oxford, OX1 3RH, UK\\
$^{6}$Naval Research Laboratory, Remote Sensing Division, 4555 Overlook Ave. SW, Washington, DC 20375, USA\\
$^{7}$Institute of Astrophysics, Universiteit Leuven, Celestijnenlaan 200 D, 3001 Leuven, Belgium\\
$^{8}$SRON Netherlands Institute for Space Research, Niels Bohrweg 4, 2333 CA, Leiden, The Netherlands\\
$^{9}$Department of Astronomy and Steward Observatory, University of Arizona, 933 N. Cherry Ave., Tucson, AZ 85721, USA\\
$^{10}$Large Binocular Telescope Observatory, University of Arizona, 933 North Cherry Avenue, Tucson, AZ 85721, USA\\
$^{11}$Minnesota Institute for Astrophysics, University of Minnesota, 116 Church Street SE, Minneapolis, MN 55455, USA\\
$^{12}$Leibniz-Institute for Astrophysics Potsdam (AIP), An der Sternwarte 16, 14482 Potsdam, Germany\\
$^{13}$Department of Astronomy and Astrophysics, University of California, Santa Cruz, 1156 High St, Santa Cruz, CA 95064, USA\\
$^{14}$Department of Astronomy, The Ohio State University, 140 W. 18th Ave., Columbus, OH 43210, USA\\
$^{15}$Center for Astrophysics \textbar~Harvard \& Smithsonian, 60 Garden Street, Cambridge, MA 02138, USA
}

\date{Accepted 2024 May 20. Received 2024 May 15; in original form 2024 February 13}

\pubyear{2024}

\begin{document}
\begin{CJK*}{UTF8}{gbsn}
\label{firstpage}
\pagerange{\pageref{firstpage}--\pageref{lastpage}}
\maketitle

\begin{abstract}
The time variability and spectra of directly imaged companions provide insight into their physical properties and atmospheric dynamics. We present follow-up R$\sim$40 spectrophotometric monitoring of red companion HD~1160~B at 2.8-4.2~$\upmu$m using the double-grating 360\textdegree{} vector Apodizing Phase Plate (dgvAPP360) coronagraph and ALES integral field spectrograph on the Large Binocular Telescope Interferometer. We use the recently developed technique of gvAPP-enabled differential spectrophotometry to produce differential light curves for HD~1160~B. We reproduce the previously reported $\sim$3.2~h periodic variability in archival data, but detect no periodic variability in new observations taken the following night with a similar 3.5\% level precision, suggesting rapid evolution in the variability of HD~1160~B. We also extract complementary spectra of HD~1160~B for each night. The two are mostly consistent, but the companion appears fainter on the second night between 3.0-3.2~$\upmu$m. Fitting models to these spectra produces different values for physical properties depending on the night considered. We find an effective temperature T\textsubscript{eff}~=~2794{\raisebox{0.5ex}{\tiny$\substack{+115 \\ -133}$}}~K on the first night, consistent with the literature, but a cooler T\textsubscript{eff}~=~2279{\raisebox{0.5ex}{\tiny$\substack{+79 \\ -157}$}}~K on the next. We estimate the mass of HD~1160~B to be 16-81~M\textsubscript{Jup}, depending on its age. We also present R~=~50,000 high-resolution optical spectroscopy of host star HD~1160~A obtained simultaneously with the PEPSI spectrograph. We reclassify its spectral type to A1\,IV-V and measure its projected rotational velocity $\varv \sin i$~=~96{\raisebox{0.5ex}{\tiny$\substack{+6 \\ -4}$}}\,km~s$^{-1}$. We thus highlight that gvAPP-enabled differential spectrophotometry can achieve repeatable few percent level precision and does not yet reach a systematic noise floor, suggesting greater precision is achievable with additional data or advanced detrending techniques.
\end{abstract}



\input{sections/01_introduction.tex}
\input{sections/02_target_properties.tex}
\input{sections/03_observations.tex}
\input{sections/04_data_reduction.tex}
\input{sections/05_generating_lcs.tex}
\input{sections/06_spectral_analysis.tex}
\input{sections/07_pepsi.tex}
\input{sections/08_discussion.tex}
\input{sections/09_conclusions.tex}
\input{sections/10_acknowledgements.tex}
\input{sections/11_data_availability.tex}




\bibliographystyle{mnras}
\bibliography{bibliography} 





\appendix



\bsp	
\label{lastpage}
\end{CJK*}
\end{document}

%% file: sections/01_introduction.tex
\section{Introduction}\label{p3_ch4_intro}
With the aid of the latest advancements in adaptive optics and coronagraphic instrumentation, the technique of direct high-contrast imaging has uncovered $\sim$30 planetary-mass companions in wide orbits around their host stars \citep[e.g.][]{2010Natur.468.1080M, 2014ApJ...780L...4B, 2017AJ....153...18B, 2018A&A...617A..44K, 2019NatAs...3..749H, 2020MNRAS.492..431B, 2020ApJ...898L..16B, 2021A&A...648A..73B, 2022NatAs...6..751C, 2023Sci...380..198C, 2023A&A...671L...5H}. Furthermore, searches for such objects have also identified a population of higher mass substellar companions up to the brown dwarf/stellar boundary \citep[e.g.][]{2010ApJ...720L..82B, 2015ApJ...811..103M, 2016ApJ...829L...4K, 2017A&A...597L...2M, 2020A&A...635A.203R, 2024A&A...684A..88R, 2022MNRAS.513.5588B, 2022ApJ...934L..18K, 2023AJ....165...39F, 2023MNRAS.522.5622L}. These brown dwarf companions are generally brighter than exoplanets and hence easier to observe, yet often appear to have similar properties to giant exoplanets, sharing similar effective temperatures, surface gravities, and weather \citep[e.g.][]{2013Sci...341.1492D, 2013MmSAI..84..955F, 2016ApJS..225...10F, 2014A&ARv..22...80H, 2016ApJ...826L..17S, 2018ApJ...858...97M, 2019MNRAS.483..480V, 2022ApJ...934..178A, 2024MNRAS.527.6624L}. Studies of brown dwarfs as exoplanet analogues may therefore also help us to understand the underlying processes in exoplanet atmospheres and to break degeneracies surrounding formation mechanisms, which may differ between the two populations despite their similarities.

While spectroscopic observations allow us to derive values for the physical parameters of brown dwarfs through comparisons of companion spectra to atmospheric models, high-cadence variability monitoring provides insight into the dynamics and structure of atmospheric features such as clouds and storms \citep[e.g.][]{2013ApJ...762...47K, 2014Natur.505..654C, 2016ApJ...825...90K, 2019ApJ...875L..15M, 2021AJ....162..179M, 2022AJ....164...65M, 2023AJ....165..181M, 2022ApJ...924...68V, 2023ApJ...944..138V, 2024arXiv240205345H, 2024AJ....167..237L}. Variability has now been detected in the light curves of numerous planetary-mass and brown dwarf companions using observations obtained with space-based telescopes \citep[e.g.][]{2016ApJ...818..176Z, 2020AJ....160...77Z, 2022AJ....164..239Z, 2019ApJ...883..181M, 2021A&A...651L...7M, 2020ApJ...893L..30B, 2020AJ....159..125L}. However, obtaining similar measurements using ground-based telescopes, which have the large diameters needed to resolve companions at close angular separations, has proven more challenging. Non-astrophysical variability induced by turbulence in Earth's atmosphere overwhelms any variability signal from the atmosphere of a faint companion. While the companion's host star would be an ideal photometric reference with which to divide out these systematics and recover its intrinsic variability, this is often obscured by the focal-plane coronagraphs used by most coronagraphic imagers \citep[e.g.][]{2018SPIE10698E..2SR}.  Nonetheless, ground-based variability studies with coronagraphic imagers have been shown to be possible using satellite spots as photometric references for the companion light curve, with which upper limits for variability have been found for the exoplanets orbiting HR~8799 \citep{2014SPIE.9147E..55W, 2022AJ....164..143W, 2015ApJ...813L..24J, 2016ApJ...820...40A, 2021MNRAS.503..743B}.

Another, more recently developed approach for exploring the variability of high-contrast companions from the ground involves using the technique of differential spectrophotometry in combination with a vector Apodizing Phase Plate (vAPP) coronagraph \citep{2023MNRAS.520.4235S}. vAPP coronagraphs offer a breakthrough in high-contrast variability searches as they provide a reliable photometric reference as a result of their intrinsic design; unlike focal-plane coronagraphs, they enable observations of high-contrast companions at close separations while maintaining a Point Spread Function (PSF) of the target star that can be used as a simultaneous photometric reference \citep{2012SPIE.8450E..0MS, 2014OExpr..2230287O, 2021ApOpt..60D..52D, 2021MNRAS.506.3224S, 2023A&A...674A.115L}. To obtain differential spectrophotometry, a vAPP can be combined with an integral field spectrograph (IFS) to spectrally disperse the light from the target. The spectra can then be recombined into white-light, reducing the impact of systematic errors in any single wavelength channel and therefore producing light curves with higher precision \citep{2023MNRAS.520.4235S}. The light curve of the companion is then divided by the light curve of the photometric reference (in this case the host star) to remove systematic variability trends shared by both objects. This concept of differential spectrophotometry was also used for the satellite spot study of \citet{2022AJ....164..143W}, and is commonplace in the field of exoplanet transmission spectroscopy \citep[e.g.][]{2018AJ....156...42D, 2020AJ....160...27D, 2023AJ....165..169D, 2019A&A...631A.169T, 2021A&A...646A..94A, 2022MNRAS.510.3236P, 2022MNRAS.515.5018P}. To demonstrate this technique, \citet{2023MNRAS.520.4235S} observed HD~1160~B, a companion with a peculiar spectrum at the brown dwarf/stellar boundary, for one night with the double-grating 360\textdegree{} vector Apodizing Phase Plate \citep[dgvAPP360,][]{2017SPIE10400E..0UD, 2020PASP..132d5002D, 2021ApOpt..60D..52D} coronagraph combined with the Arizona Lenslets for Exoplanet Spectroscopy (ALES) IFS on the Large Binocular Telescope. They detected significant sinusoidal variability in the differential white-light curve of this companion with a semi-amplitude of 8.8\% and a period of $\sim$3.24 hours. Furthermore, they obtained a 3.7\% precision in bins of 18~minutes, after a multiple linear regression approach was applied to the differential white-light curve to remove residual systematics arising from non-astrophysical sources such as airmass and detector position. This study found no evidence of having reached a systematic noise floor in their single epoch of observations, indicating that the data was not systematic-limited and that additional data could further improve the sensitivity to variability.

In this work, we further monitor and characterize HD~1160~B and its host star HD~1160~A using additional observations obtained with the Large Binocular Telescope (LBT). This includes a further night of variability monitoring of HD~1160~B with ALES+dgvAPP360, which we also use to produce its complementary spectral characterization using the 2.8-4.2~$\upmu$m spectra from both epochs. We separately characterize the host star HD~1160~A using data obtained simultaneously with the PEPSI high resolution spectrograph. The differential spectrophotometry observations with ALES+dgvAPP360 also allow us to test the repeatability of the light curve precision achieved by \citet{2023MNRAS.520.4235S} in their pilot study. In Section~\ref{p3_ch4_target_properties}, we review the properties of the HD~1160 system. Our observations of this system are then described in Section~\ref{p3_ch4_obs}, and in Section~\ref{p3_ch4_data_red} we describe the methods used to reduce the data from each instrument and extract photometry of the targets. In Section~\ref{p3_ch4_diff_lcs_section}, we investigate the variability of HD~1160~B by using the ALES+dgvAPP360 data to produce differential spectrophotometric light curves. We also use these data in Section~\ref{p3_ch4_spectral_results} to produce and study the spectrum of HD~1160~B. The data obtained with PEPSI are analysed in Section \ref{p3_ch4_PEPSI_method_results}, in which we explore the spectrum of HD~1160~A. The results found in each of these three sections are then discussed in Section \ref{p3_ch4_discussion}. Finally, we summarise the conclusions of the paper in Section \ref{p3_ch4_conclusions}.

%% file: sections/02_target_properties.tex
\section{Target Properties}\label{p3_ch4_target_properties}
The HD~1160 system is located at a distance of 120.7$\pm$0.5~pc \citep[Gaia Data Release 3,][]{2016A&A...595A...1G, 2023A&A...674A...1G}. In Table~\ref{table:p3_ch4_prim_parameters}, we summarize literature values for key properties of the stellar primary component HD~1160~A, for which \cite{1999mctd.book.....H} assigned a spectral type of A0\,V using photographic plates on the 0.61-m Curtis Schmidt telescope at the Cerro Tololo Inter-American Observatory (CTIO). Using observations from the Transiting Exoplanet Survey Satellite (TESS) mission, \citet{2023MNRAS.520.4235S} found that HD~1160~A is non-variable in the optical at the 0.03\% level, and Spitzer observations by \citet{2006ApJ...653..675S} found no infrared excess, suggesting that there is not significant warm circumstellar dust present. \citet{2012ApJ...750...53N} identified two comoving companions to HD~1160~A at separations of $\sim$80~au ($\sim$0.78$\arcsec$) and $\sim$530~au ($\sim$5.15$\arcsec$), known as HD~1160~B and C, respectively, during the Gemini Near-Infrared Coronagraphic Imager (NICI) Planet-Finding Campaign \citep{2010SPIE.7736E..1KL}. HD~1160~B has a contrast of $\Delta L^{\prime}=6.35\pm0.12$ mag relative to the $L^{\prime}=7.055\pm0.014$ mag of HD~1160~A, and its orbit is almost edge-on, with an 
inclination angle of 92$^{+8.7}_{-9.3}$\textdegree \citep{2003MNRAS.345..144L, 2012ApJ...750...53N, 2020AJ....159...63B}. The wide angular separation of HD~1160~C places it beyond the fields of view of the data sets in this paper.

\citet{2012ApJ...750...53N} found HD~1160~B to be an L0$\pm$2 brown dwarf based on their near-infrared photometry, and that their near-infrared spectrum of HD~1160~C best matches that of an M3.5$\pm$0.5 low-mass star. They found that both companions are redder than similar objects, which combined with an apparent underluminosity of HD~1160~A suggests a young age of 50{\raisebox{0.5ex}{\tiny$\substack{+50 \\ -40}$}}~Myr. Combining this age range with the luminosity of HD~1160~B, they derived a value for its mass of 33{\raisebox{0.5ex}{\tiny$\substack{+12 \\ -9}$}}~M\textsubscript{Jup}.

However, using observations from the Spectro-Polarimetric High-contrast imager for Exoplanets REsearch \citep[SPHERE,][]{2019A&A...631A.155B} instrument on the Very Large Telescope (VLT), \citet{2016A&A...587A..56M} concluded that the 1.0-1.6~$\upmu$m spectrum of HD~1160~B best matched that of a M6.0{\raisebox{0.5ex}{\tiny$\substack{+1.0 \\ -0.5}$}} dwarf. Unlike \citet{2012ApJ...750...53N}, they did not find unusually red colours for either companion. They also found higher estimates for its mass; 79{\raisebox{0.5ex}{\tiny$\substack{+65 \\ -40}$}}~M\textsubscript{Jup} based on its luminosity and 107{\raisebox{0.5ex}{\tiny$\substack{+59 \\ -38}$}}~M\textsubscript{Jup} based on its effective temperature. The wide range of possible masses is driven by the uncertain age used, 100{\raisebox{0.5ex}{\tiny$\substack{+200 \\ -70}$}}~Myr, which was chosen due to the lack of reliable age indicators with the upper limit given by the 300~Myr predicted main-sequence lifetime of an A0 star \citep{2000A&A...358..593S}.

\citet{2017ApJ...834..162G} also observed the HD~1160 system, using the Subaru Coronagraphic Extreme Adaptive Optics \citep[SCExAO][]{2015PASP..127..890J} instrument and the Gemini Planet Imager \citep[GPI, ][]{2014PNAS..11112661M}. They too found that HD~1160~B has typical colours for a mid-M dwarf and assign it a spectral type of M5.5{\raisebox{0.5ex}{\tiny$\substack{+1.0 \\ -0.5}$}}, in good agreement with \citet{2016A&A...587A..56M}, and rule out earlier spectral types. Considering a range of different evolutionary models, they report two different possible system ages; 20-125~Myr if HD~1160~A is considered alone, and 80-125~Myr if HD~1160~A, B, and C are considered jointly. These lead to mass values for HD~1160~B of 35-90~M\textsubscript{Jup} and 70-90~M\textsubscript{Jup}, respectively. However, they note that the derived mass of HD~1160~B is highly dependent on its surface gravity and age. \citet{2017ApJ...834..162G} further found that HD~1160~B likely has approximately solar metallicity, which is consistent with almost all systems in the solar neighbourhood \citep[e.g.][]{2002A&A...389..871D}.

Based on its Gaia kinematics, \citet{2019AJ....158...77C} found that the HD~1160 system could be part of the Pisces-Eridanus stellar stream, indicating an age on the order of $\sim$120-135~Myr if this were to be confirmed \citep{2019A&A...622L..13M, 2020A&A...638A...9R}.

The most recent spectral characterization of HD~1160~B was carried out by \citet{2020MNRAS.495.4279M}, who again observed the system with SPHERE and found it to have a peculiar spectrum that is not well matched by any spectra in current spectral libraries, but concluded a spectral type of M5-M7 based on the best fits to individual spectral bands. They propose that this peculiarity could be explained by the presence of dust in its photosphere, or if it has a young age and is not yet fully matured. By fitting the spectrum of HD~1160~B with atmospheric models and considering alkali lines that become weaker at lower surface gravities, \citet{2020MNRAS.495.4279M} found a low surface gravity of log(g) = 3.5-4.0 dex. This suggests that HD~1160~B may actually have a young age of 10-20~Myr, and a mass of $\sim$20~M\textsubscript{Jup}, in contrast to previous results. However, they noted that they cannot rule out older ages.

While the studies above explored the spectrum of HD~1160~B, it was also the target of a variability monitoring search by \citet{2023MNRAS.520.4235S}. As described in Section~\ref{p3_ch4_intro}, they found 8.8\% semi-amplitude variability with a period of $\sim$3.24 hours in the differential L-band white-light curve of HD~1160~B during a pilot study combining the technique of differential spectrophotometry with the dgvAPP360 coronagraph. They attribute this variability to heterogeneous features in the atmosphere of the companion, such as clouds or cool star spots, but conclude that additional data is needed to confirm its periodicity and establish its physical explanation.
\begin{flushleft}
\begin{table}
\caption{Properties of host star HD~1160~A.}
\begin{tabular}{p{0.5\columnwidth}p{0.2\columnwidth}p{0.1\columnwidth}}
\hline
Property&Value&Ref.\\
\hline
Right Ascension (J2000, hh:mm:ss.ss)&00:15:57.32&(1)\\
Declination (J2000, dd:mm:ss.ss)&+04:15:03.77&(1)\\
RA proper motion (mas yr$^{-1}$)&20.150$\pm$0.040&(1)\\
Dec. proper motion (mas yr$^{-1}$)&-14.903$\pm$0.034&(1)\\
Parallax (mas)&8.2721$\pm$0.0355&(1)\\
Radial velocity (km s$^{-1}$)&13.5$\pm$0.5&(1)\\
Distance (pc)&120.7$\pm$0.5&(1)\\
Extinction A\textsubscript{V} (mag)&0.16&(1)\\
Spectral Type&A0\,V&(2)\\
&A1\,IV-V&(3)\\
Mass (M$_{\odot}$)&$\sim$2.2&(4)\\
T\textsubscript{eff} (K)&9011$\pm$85&(5)\\
&9200{\raisebox{0.5ex}{\tiny$\substack{+200 \\ -100}$}}&(3)\\
log(g) (dex)&$\sim$4.5&(6)\\
&3.5{\raisebox{0.5ex}{\tiny$\substack{+0.5 \\ -0.3}$}}&(3)\\
$\varv \sin i$ (km~s$^{-1}$)&96{\raisebox{0.5ex}{\tiny$\substack{+6 \\ -4}$}}&(3)\\
log(L/L$_{\odot}$)&1.12$\pm$0.07&(5)\\
{[Fe/H]}&$\sim$solar&(6)\\
V (mag)&7.119$\pm$0.010&(7)\\
{\it Gaia} G (mag)&7.1248$\pm$0.0004&(1)\\
J (mag)&6.983$\pm$0.020&(8)\\
H (mag)&7.013$\pm$0.023&(8)\\
K (mag)&7.040$\pm$0.029&(8)\\
L$^{\prime}$ (mag)&7.055$\pm$0.014&(9)\\
M$^{\prime}$ (mag)&7.04$\pm$0.02&(9)\\
\hline
System age (Myr)&50{\raisebox{0.5ex}{\tiny$\substack{+50 \\ -40}$}}&(4)\\
&100{\raisebox{0.5ex}{\tiny$\substack{+200 \\ -70}$}}&(10)\\
&20-125&(5)\\
&$\sim$120&(11)\\
&10-20&(6)\\
\hline
\end{tabular}
\textbf{References:} (1) \citet{2016A&A...595A...1G, 2023A&A...674A...1G}; (2) \citet{1999mctd.book.....H}; (3) This work; (4) \citet{2012ApJ...750...53N}; (5) \citet{2017ApJ...834..162G}; (6) \citet{2020MNRAS.495.4279M}; (7) Tycho-2 \citep{2000A&A...355L..27H, 2000A&A...357..367H}; (8) 2MASS \citep{2003yCat.2246....0C, 2006AJ....131.1163S}; (9) \citet{2003MNRAS.345..144L}; (10) \citet{2016A&A...587A..56M}; (11) \citet{2019AJ....158...77C}
\label{table:p3_ch4_prim_parameters}
\end{table}
\end{flushleft}

%% file: sections/03_observations.tex
\section{Observations}\label{p3_ch4_obs}
We observed the HD~1160 system on the nights of 2020 September 25 (03:27:31 - 11:16:14 UT) and 2020 September 26 (03:20:16 - 10:46:09 UT) using the 2~x~8.4-m Large Binocular Telescope (LBT) at the Mount Graham International Observatory, Arizona. On the left-hand side aperture of the LBT, we used the dgvAPP360 coronagraph (see Section~\ref{p3_ch4_intro}) in combination with the Arizona Lenslets for Exoplanet Spectroscopy (ALES) IFS \citep{2015SPIE.9605E..1DS, 2018SPIE10702E..3LH, 2018SPIE10702E..3FS}. ALES is located in the focal plane of the LBT mid-infrared camera \citep[LMIRcam,][]{2008SPIE.7013E..3AW, 2010SPIE.7735E..3HS, 2012SPIE.8446E..4FL} and mounted inside the LBT Interferometer \citep[LBTI,][]{2015SPIE.9605E..1GD, 2016SPIE.9907E..04H, 2020SPIE11446E..07E}, which uses the LBTI adaptive optics (AO) system to provide a Strehl ratio up to 90\% at 4~$\upmu$m \citep{2012SPIE.8445E..0UH, 2014SPIE.9148E..03B, 2016SPIE.9909E..3VP, 2021arXiv210107091P}. These observations were obtained using the ALES L-band prism, providing R$\sim$40 spectroscopy over a 2.8-4.2~$\upmu$m wavelength range simultaneously, with a 2.2$\arcsec$x 2.2$\arcsec$ field of view and plate scale of $\sim$35~mas~spaxel$^{-1}$ \citep{2018SPIE10702E..0CS}.

The first night of these LBT/ALES+dgvAPP360 observations has previously been described by \citet{2023MNRAS.520.4235S}. On the second night, we obtained 2000 ALES frames with 5.4~s of integration time per frame, ensuring that the stellar PSF remained unsaturated in each frame. The total time on-target was therefore 10800~s or 3~h \citep[compared to $\sim$3.32~h on the first night, over 2210 frames of the same integration time,][]{2023MNRAS.520.4235S}. However, this on-target integration time is spread out over $\sim$7.43~h due to time spent on nodding, wavelength calibrations, and readout time. When we combine both nights of data, the total on-target integration time is 22734~s (6.32~h) over a timescale of 112718~s ($\sim$31.31~h, $\sim$1.30~days). To enable background subtraction, both nights used an on/off nodding pattern, switching position every 10~min except when interrupted by an open AO loop or to take wavelength calibrations. We also obtained 6 wavelength calibrations at irregular intervals throughout the night, and dark frames at the end of the night with the same exposure time as the science and calibration frames. At a separation of $\sim$0.78$\arcsec$, HD~1160~B remained in the coronagraphic dark hole of the dgvAPP360 at all wavelengths throughout the observations, while HD~1160~C was beyond the 2.2$\arcsec\times$~2.2$\arcsec$ field of view of ALES at $\sim$5.1$\arcsec$.

On the right-hand side LBT aperture, we used the Potsdam Echelle Polarimetric and Spectroscopic Instrument (PEPSI), a fiber-fed white-pupil echelle spectrograph \citep{2015AN....336..324S, 2018SPIE10702E..12S}. We obtained high resolution (R~=~50,000) optical spectra on both nights using the 300~$\upmu$m diameter PEPSI fiber, which operates over a wavelength range of 383-907nm. This fiber has a diameter of 2.25$\arcsec$ which encompasses the angular separation of HD~1160~B from HD~1160~A ($\sim$0.78$\arcsec$), so the obtained PEPSI spectra are combined spectra of both objects. HD~1160~C was located outside of the fiber at a separation of $\sim$5.1$\arcsec$. Data were obtained with the first three and the sixth PEPSI cross dispersers (CDs), which cover wavelength ranges of 383.7-426.5~nm, 426.5-480.0~nm, 480.0-544.1~nm, and 741.9-906.7~nm, respectively, but not with the fourth and fifth CDs. We observed using two CDs at any given time; the sixth CD was always in use, and was paired with one of the other three on a rotating cycle. The total on-target integration times obtained with each CD were 14713~seconds, 14761~seconds, 14666~seconds, and 44723~seconds for CDs 1, 2, 3, and 6, respectively.

On the second night, no time was lost to weather and the observing conditions were stable with no cloud cover. The seeing ranged from 0.7-1.5$\arcsec$. LBTI is by design always pupil-stabilized, with no instrument derotator. This means that all data are inherently obtained in pupil-stabilized mode such that the companion position rotates in the field of view with the sky. The centre of this rotation was HD~1160~A as this was used as the AO reference star. The total field rotation over the course of the night was 108.2\textdegree{}. This is comparable to the 109.7\textdegree{} of field rotation on the first night, on which the observing conditions were similarly clear with a seeing of 0.7-1.4$\arcsec$ \citep{2023MNRAS.520.4235S}. These observations were successfully scheduled during suitable nights as part of LBTI's queue scheduling, which was critical for obtaining high-quality data on two consecutive nights.

%% file: sections/04_data_reduction.tex
\section{Data reduction and spectral extraction} \label{p3_ch4_data_red}
\subsection{LBT/ALES+dgvAPP360 data processing}\label{p3_ch4_nales_processing}
Our goal is to use the LBT/ALES+dgvAPP360 observations to characterize HD~1160~B by measuring both its spectrum and its time variability. We therefore need to construct a flux-calibrated spectrum of the companion by summing the observations in the time dimension, and a `white-light' curve of the companion by summing the observations in the wavelength dimension.

Several data processing steps are required to convert the raw ALES data from 2D grids of micro-spectra on the detector into 3D image cubes of x,y-position and wavelength and prepare them for our analyses \citep{2019AJ....157..244B, 2022AJ....163..217D, 2022SPIE12184E..42S}. The data from the first night of LBT/ALES+dgvAPP360 observations was previously processed (for a time variability study only) by \citet{2023MNRAS.520.4235S}. We reprocessed this first night of data here following the same method as \citet{2023MNRAS.520.4235S}, and also used this approach for the data from the second night to ensure consistency between the two epochs. We briefly summarise the steps in this process here. We first used the sky frames from the off-source nod position to subtract the background in each frame, before extracting the micro-spectra into 3D cubes through weighted optimal extraction, where the extraction weights were defined by the wavelength-averaged spatial profiles of the micro-spectra in the sky frames \citep{1986PASP...98..609H, 2018SPIE10702E..2QB, 2020AJ....160..262S}. Next, the data were wavelength calibrated using four narrow-band filters operating upstream of the ALES optics. Each of these filters produced a single-wavelength spot on the LMIRcam detector. We performed the wavelength calibration for the 63$\times$67 micro-spectra in the ALES grid by fitting the positions of these four spots with a second-order polynomial to derive the necessary wavelength solution \citep[][]{2018SPIE10702E..3FS, 2022SPIE12184E..42S}. This process produced 3D wavelength-calibrated data cubes of $x$- and $y$-position (63$\times$67 pixels), and wavelength $\lambda$, with 100 wavelength channels spanning the 2.8-4.2~$\upmu$m wavelength range of ALES.

\begin{figure}
\centering
  \textcolor{white}{\frame{\includegraphics[width=0.95\columnwidth]{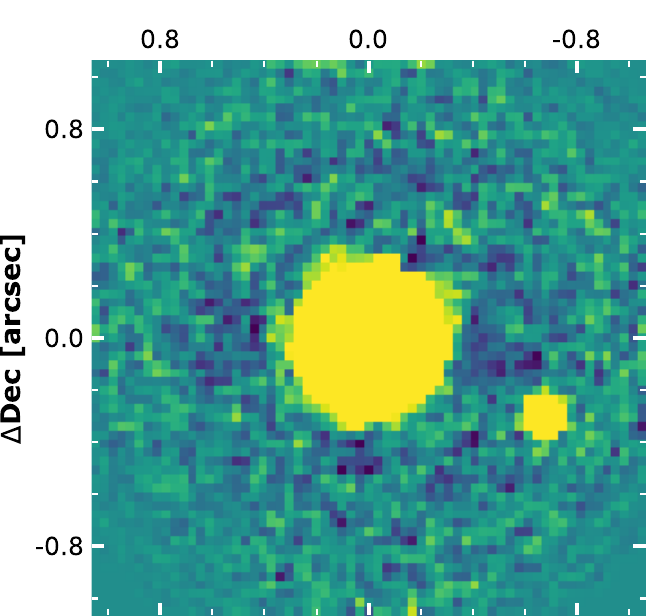}}}\\
  \textcolor{white}{\frame{\includegraphics[width=0.95\columnwidth]{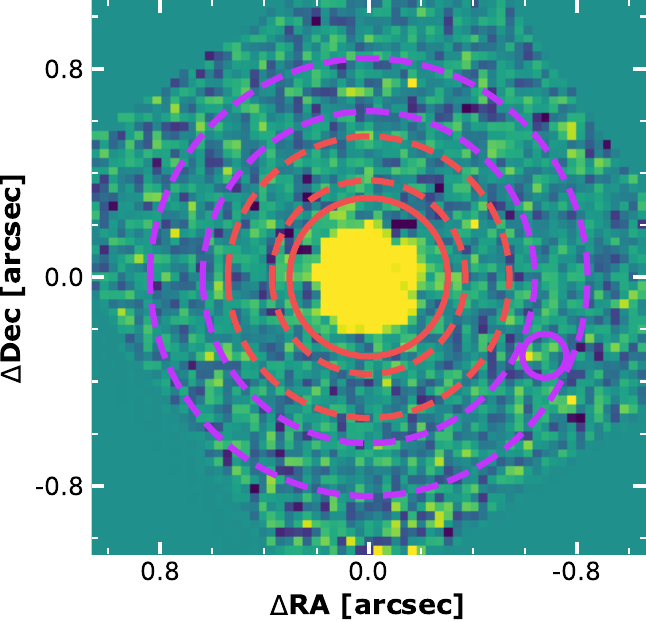}}}%
\caption{Top panel: the final LBT/ALES+dgvAPP360 image of the HD~1160 system from the second night, produced by taking the median combination of all frames in the 3.59-3.99~$\upmu$m range over both time and wavelength. This image covers a total integration time of 10800~s (3~h). Bottom panel: A single frame of data from the 3.69~$\upmu$m wavelength channel, overplotted with the apertures and annuli used to obtain flux and background measurements for the host star HD~1160~A (in orange) and companion HD~1160~B (in purple). The dashed lines indicate the background annuli. Each image uses a different arbitrary logarithmic colour scale, and both are north-aligned, where north is up and east is to the left.}
\label{fig:p3_ch4_fits_apertures}
\end{figure}

Continuing to follow the data reduction method of \citet{2023MNRAS.520.4235S}, we removed 8 frames from the first night that were unsuitable as the AO loop opened during the exposure. No frames were removed from the dataset from the second night. We then performed a bad pixel correction for each frame before applying a flat-field correction. The flat frame used for this was created from images obtained in the off-source nod position. ALES images also contain systematic time-varying row and column discontinuities caused by the intersection of the ALES micro-spectra with the channels of the LMIRcam detector \citep{2022AJ....163..217D}. To correct for the column discontinuities, we first masked HD~1160~A and B in each frame before fitting third-order polynomials to each column. These values were then subtracted and the process was repeated for each row to remove the row discontinuities. The frames were then shifted using a spline interpolation to centre the star in each frame and derotated using their parallactic angles to align them to north. A final image from the second night, obtained by median-combining every frame in the 3.59-3.99~$\upmu$m range in both time and wavelength, is shown in the top panel of Figure~\ref{fig:p3_ch4_fits_apertures}. Both HD~1160~A and B are clearly visible.

As the data were obtained in pupil-stabilized mode, we could have applied post-processing algorithms reliant on angular diversity \citep[e.g. Angular Differential Imaging,][]{2006ApJ...641..556M} to further remove quasistatic speckle noise and increase the S/N of the targets. However, we chose not to do this so that we could make use of the stellar PSF provided by the dgvAPP360 as a simultaneous photometric reference when characterizing the variability of HD~1160~B in Section \ref{p3_ch4_diff_lcs_section}. If we had applied an ADI-based algorithm the stellar PSF would have been removed, meaning there would be no photometric reference with which to divide out time-varying systematics from the companion flux.

\subsubsection{Photometric extraction}\label{p3_ch4_section:aper_phot}
Once the data had been fully processed to correct for the systematic discontinuities we extracted simultaneous aperture photometry of HD~1160~A and B, again following the approach of \citet{2023MNRAS.520.4235S}. Although some of the 100 ALES wavelength channels are not suitable for analysis (see Sections \ref{p3_ch4_4_wl_channel_select} and \ref{p3_ch4_spectral_extraction_fluxcalib}), we nonetheless performed this step for every frame in each of the 100 channels to allow a selection to be carried out at a later stage in the process. To do this, we extracted photometry in circular apertures with radii of 9 pixels (3.1~$\lambda$/D) for HD~1160~A and 2.5 pixels (0.9~$\lambda$/D) for HD~1160~B. The background flux was near zero following the removal of the row and column discontinuities in the previous section. However, we nonetheless estimated the residual background at the locations of the star and companion such that we could correct our flux measurements for any remaining offset. The background flux at the location of HD~1160~A was estimated by extracting photometry in a circular annulus with inner and outer radii of 11 and 16 pixels, respectively. The drift of the star combined with the rotation of the field over the course of the night means that HD~1160~B was close to the edge of the field of view in some frames, meaning that we could not use a similar annulus to estimate the background at its location. Instead, we did this by masking HD~1160~B and then extracting photometry in another annulus centred on the star, this time with a 6-pixel width around the radial separation of HD~1160~B \citep{2021MNRAS.503..743B, 2023MNRAS.520.4235S}. We show these apertures and annuli overplotted on a single frame of data in the bottom panel of Figure~\ref{fig:p3_ch4_fits_apertures}. We then corrected our aperture photometry of the star and companion by subtracting the mean counts per pixel in the corresponding annulus multiplied by the area of the respective aperture.

This extracted spectrophotometry of HD~1160~A and B is used to investigate the time variability of HD~1160~B in Section~\ref{p3_ch4_diff_lcs_section} and its spectrum in Section~\ref{p3_ch4_spectral_results}.

\subsection{PEPSI data processing}\label{p3_ch4_pepsi_processing}
In this paper we aim to use the LBT/PEPSI observations to characterize the physical properties of host star HD~1160~A. The PEPSI data were reduced using the Spectroscopic Data Systems for PEPSI (SDS4PEPSI) generic package written in C++ under Linux and based on the 4A package for processing data from the SOFIN spectrograph on the Nordic Optical Telescope \citep{2000PhDT..........I, 2018A&A...612A..44S, 2018A&A...612A..45S, 2022MNRAS.513.1544K}. SDS4PEPSI applied a fully automated set of standardised data reduction steps to the raw data, including CCD bias removal, photon noise estimation, flat-field correction, and scattered light subtraction. It then performed a weighted optimal extraction of the spectral orders to maximise the S/N of the target, and then performed wavelength calibration. The spectra were then normalised to the continuum by fitting the extracted spectral orders with a 2D smoothing spline on a regular grid of CCD pixels and echelle order numbers, and then shifted to the stellar rest frame. Each of these steps carried out by SDS4PEPSI is described in full detail by \citet{2018A&A...612A..44S}. Finally, we combined the spectra from all of the exposures obtained with a given CD by interpolating them to the same wavelengths and summing them according to their weights, where the weights are defined as the inverse of the noise.

%% file: sections/05_generating_lcs.tex
\section{Analysing the variability of HD~1160~B}\label{p3_ch4_diff_lcs_section}
In this section we use the aperture photometry of HD~1160~A and B obtained in Section~\ref{p3_ch4_section:aper_phot} to explore the time variability of HD~1160~B via the technique of differential spectrophotometry. This method applied using the dgvAPP360 coronagraph was first described by \citet{2023MNRAS.520.4235S}. While we are interested in the intrinsic variability arising from the atmosphere of the companion, the raw flux that we obtained through aperture photometry is inherently polluted by additional variability caused by Earth's atmosphere and systematics originating from the instrumentation. This unwanted variability can be mitigated using an independent, simultaneous photometric reference, but this is generally problematic for ground-based variability studies of high-contrast companions, as field stars are rarely available and focal-plane coronagraphs block the host star in order to allow companions to be detected \citep[e.g.][]{2012SPIE.8442E..04M, 2018SPIE10698E..2SR}. The dgvAPP360 coronagraph uniquely enables host stars and their companions to be imaged simultaneously, thus we can use the simultaneous aperture photometry of HD~1160~A as a photometric reference to remove variability arising from non-astrophysical sources external to HD~1160~B \citep{2020PASP..132d5002D, 2021ApOpt..60D..52D, 2023MNRAS.520.4235S}. As HD~1160~A does not fit into any known category of variable star and \citet{2023MNRAS.520.4235S} previously found no evidence for variability in HD~1160~A above the 0.03\% level in TESS observations over a 51 day baseline, we proceed with the assumption that HD~1160~A does not have intrinsic variations of its own at longer wavelengths.

\subsection{ALES wavelength channel selection}\label{p3_ch4_4_wl_channel_select}
The first step in the process of making a differential white-light curve for HD~1160 was to select which wavelength channels should be included. A benefit of the spectrophotometric approach is that channels with low target S/N or issues that could introduce false variability signals can be excluded, allowing the light curve precision to be maximised. Our data cubes consist of 100 wavelength channels ranging from 2.8-4.2~$\upmu$m. However, wavelength channels at the start and end of this range are unsuitable for analysis as the photometry is contaminated by flux from the neighbouring spaxel in the dispersion direction, and those in the $\sim$3.25-3.5~$\upmu$m wavelength range are significantly impacted by absorption caused by the glue molecules in the dgvAPP360 \citep{2017ApJ...834..175O, 2021ApOpt..60D..52D, 2022AJ....163..217D}. \citet{2023MNRAS.520.4235S} selected the 30 wavelength channels in the 3.59-3.99~$\upmu$m range that had a high target S/N for inclusion in their time variability analysis of the first night of data. We therefore chose to use these same channels for our variability analysis in this paper such that we could directly compare the light curves from each night of data. We discuss the spectral data obtained on each night over the full 2.8-4.2~$\upmu$m wavelength range covered by ALES in Section~\ref{p3_ch4_spectral_extraction_fluxcalib}.

\subsection{Differential spectrophotometric light curves}\label{p3_ch4_4_diff_lcs_binning}
\begin{figure*}
	\includegraphics[width=\textwidth]{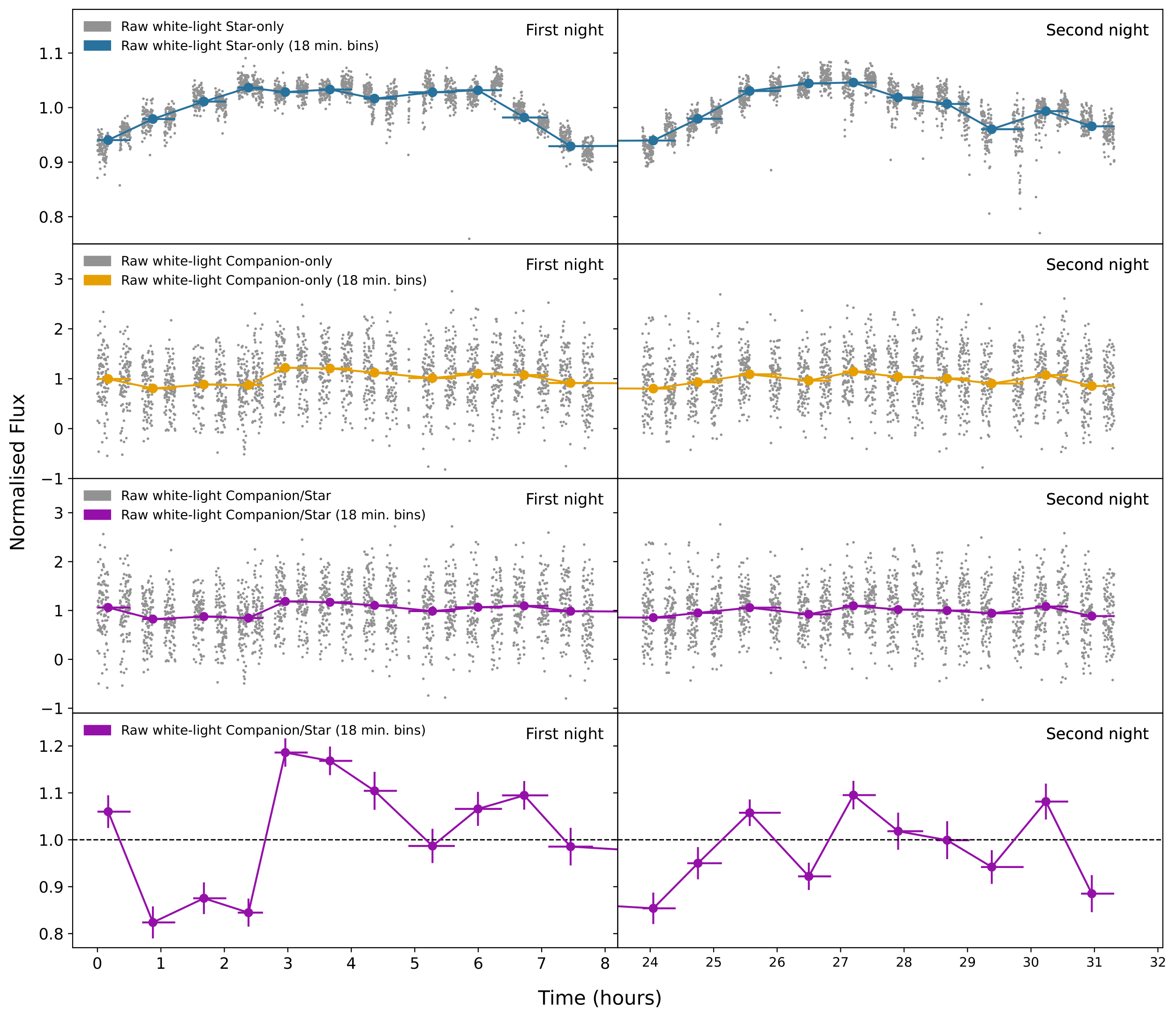}
    \caption{Top two rows: the raw white-light fluxes of host star HD~1160~A and companion HD~1160~B from both nights are plotted in grey, and binned to 18 minutes of integration time per bin in blue and yellow, respectively. The time series were normalised over the full time series covering both epochs, and consist of the data in the 3.59-3.99~$\upmu$m wavelength range. The data from the first night is reproduced from \citet{2023MNRAS.520.4235S}, but the normalisation is different here. Bottom two rows: the raw differential white-light curve obtained by dividing the unnormalised, unbinned companion flux by that of the star, shown unbinned in grey and binned in purple. The bottom row shows a zoomed-in view of the binned version. This division removes variability shared by both the star and the companion from the companion flux, leaving a differential light-curve containing only variability arising from the companion's atmosphere and non-shared systematics.}
    \label{fig:p3_ch4_raw_lc_both}
\end{figure*}

We produced our differential white-light curve of HD~1160~B following the technique presented by \citet{2023MNRAS.520.4235S}. First, we separately prepared white-light time series for HD~1160~A and HD~1160~B. We did this by taking the median combination of the photometry for each object over the 30 wavelength channels chosen in the previous section, thereby obtaining a single white-light measurement for each object at each time. These are shown in grey in the top two rows of Figure~\ref{fig:p3_ch4_raw_lc_both}, and binned to 18 minutes of integration time per bin in blue and yellow for the host star and companion, respectively. The time series shown here are plotted on the same axes and were normalised over the full sequence, including both epochs, to allow comparison between each night. Aside from the change in the normalisation, the data points on the first night are identical to those of \citet{2023MNRAS.520.4235S}. Thus, at this point, our analysis of the first night of observations deviates slightly from that presented by \citet{2023MNRAS.520.4235S}.

The gaps in integration time in the unbinned data are due to the on/off nodding pattern used when observing, and the unequal $x$-uncertainties on the binned datapoints occur where the bins overlap multiple nods. To produce a differential white-light curve, we then divided the unbinned, unnormalised flux of HD~1160~B by that of HD~1160~A. This raw differential light curve is plotted unbinned in grey, and binned in purple, in the third row of Figure~\ref{fig:p3_ch4_raw_lc_both}. We also plot a closer view of the same binned light curve in the fourth row. We calculated the errors on the binned fluxes by taking the 1.48 $\times$ median absolute deviation (MAD) of the fluxes in each time bin, then dividing these values by $\sqrt{N-1}$, where $N$ is the number of frames per bin. Dividing the two time series in this way has the effect of removing most of the variability due to shared systematics arising from the instrumentation or telluric effects. HD~1160~A is known to be non-varying to at least the 0.03$\%$ level \citep{2023MNRAS.520.4235S}, so remaining variability trends in this differential light curve are therefore only those arising from HD~1160~B itself and from any contaminating systematics that are not shared by the star and the companion.

\subsection{Detrending}\label{p3_ch4_linreg_detrend}
\begin{figure*}
	\includegraphics[width=\textwidth]{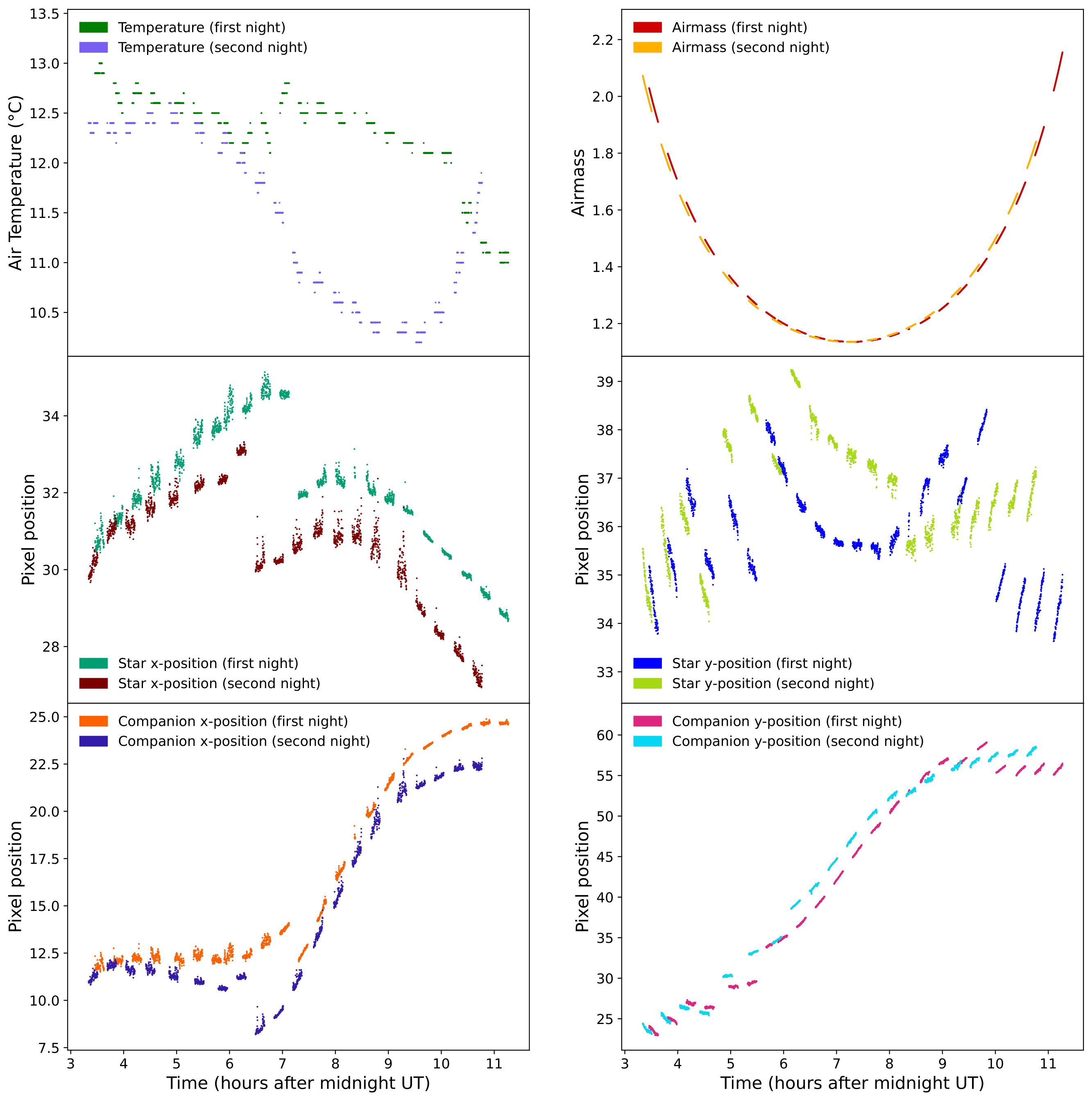}
    \caption{The six decorrelation parameters used in the linear regression to detrend the differential white-light curve of HD~1160~B, shown for both nights. To allow the trends at each epoch to be overplotted and compared, 24~hours has been removed from the $x$-axis for the second night. As with the time series photometry, the gaps in the data arise from the use of the on/off nodding pattern. The top two panels show the air temperature in \textcelsius~and the airmass as a function of time. The remaining four panels show the x- and y-positions (in pixels) of host star HD~1160~A and companion HD~1160~B in the original 3D image cubes (i.e. before spatial and rotational alignment) as a function of time. The large jumps in these positions were caused by manual offsets applied to maintain the central location of HD~1160~A within the small field of view, and the slowly varying trends arise from lenslet array flexure as the telescope rotates. For HD~1160~B, the rotation of the field of view itself (109.7\textdegree{} and 108.2\textdegree{} for the first and second nights, respectively) induces an additional component to its positional trends.}
    \label{fig:p3_ch4_linreg_parameters}
\end{figure*}

In this section we attempt to fit and remove several residual (i.e. non-shared) systematic trends from the differential light curve, with the aim of producing a detrended differential light curve containing only the intrinsic variability of HD~1160~B. These residual systematics are likely due to differences in brightness, colour, or position of the companion and host star, and can arise from both instrumental and telluric sources \citep[e.g.][]{2005AN....326..134B, 2006MNRAS.373..231P, 2012MNRAS.419.2683G, 2013A&A...550A..54D, 2018AJ....156...42D, 2022MNRAS.510.3236P, 2022MNRAS.515.5018P}. Here, we applied a multiple linear regression including six different possible sources of systematics as decorrelation parameters. These parameters are shown plotted against time in hours after midnight, for each epoch, in Figure~\ref{fig:p3_ch4_linreg_parameters}. \citet{2023MNRAS.520.4235S} found that airmass and external air temperature were the parameters that were the most correlated with the differential light curve from the first night alone, so we chose to include both of these again here. We also again included the x- and y- pixel positions of HD~1160~A and B in the original images, before centering and rotational alignment were carried out. These parameters probe any remaining systematics arising from the response of the detector or other instrumental effects. The sharp jumps in position seen in Figure~\ref{fig:p3_ch4_linreg_parameters} arise from manual positional offsets performed during the observing sequence to ensure the star did not drift too far from the centre of the small field of view. We further considered including wind speed and wind direction, but \citet{2023MNRAS.520.4235S} found that wind speed and wind direction were not significantly correlated with the trends in the light curve from the first night. We found that this was also the case for the second night, so chose not include these as parameters in the linear regression here. Thus, in addition to the different normalisation applied in Section~\ref{p3_ch4_4_diff_lcs_binning}, this is the other point at which our analysis of the first night of observations is slightly different to that of \citet{2023MNRAS.520.4235S}.

\begin{figure*}
	\includegraphics[width=\textwidth]{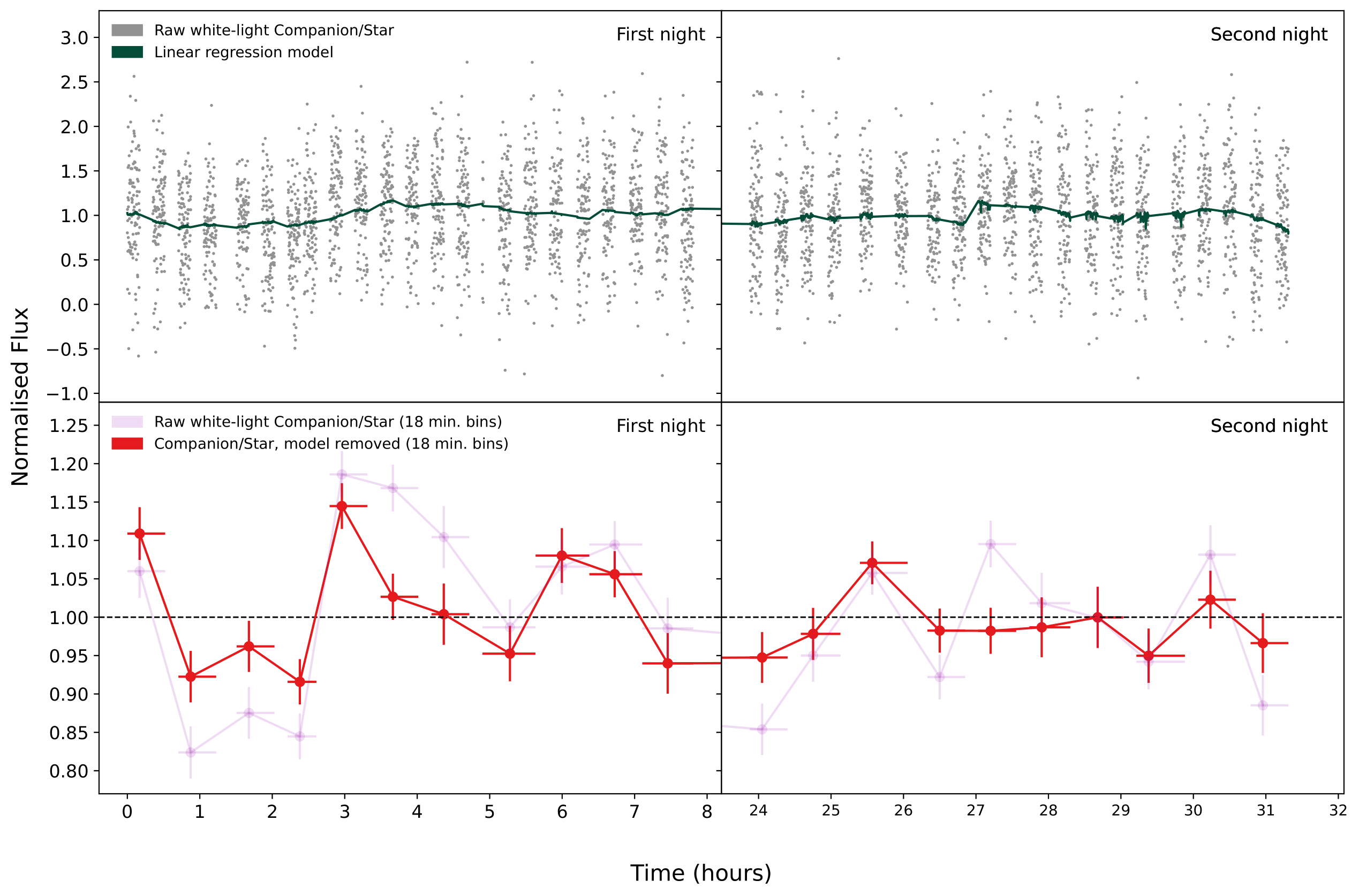}
    \caption{Top panel: the model, in dark green, produced when a multiple linear regression is applied to the raw white-light differential light curves from each night using the decorrelation parameters from Figure~\ref{fig:p3_ch4_linreg_parameters}. The corresponding coefficients and intercept of each model are given in Table~\ref{table:p3_ch4_lin_reg_coeffs}. The raw differential white-light curve is shown in grey, reproduced from the third panel of Figure~\ref{fig:p3_ch4_raw_lc_both}. Bottom panel: the detrended differential white-light curve, in red and binned to 18 minutes of integration time per bin, obtained by dividing the raw differential light curve by the linear regression models above. The binned version of the raw differential white-light curve from the bottom panel of Figure~\ref{fig:p3_ch4_raw_lc_both} is also shown for comparison in purple.}
    \label{fig:p3_ch4_detrended_lc_both}
\end{figure*}

\begin{figure*}
    \centering
  \textcolor{white}{\frame{\includegraphics[width=0.5\textwidth]{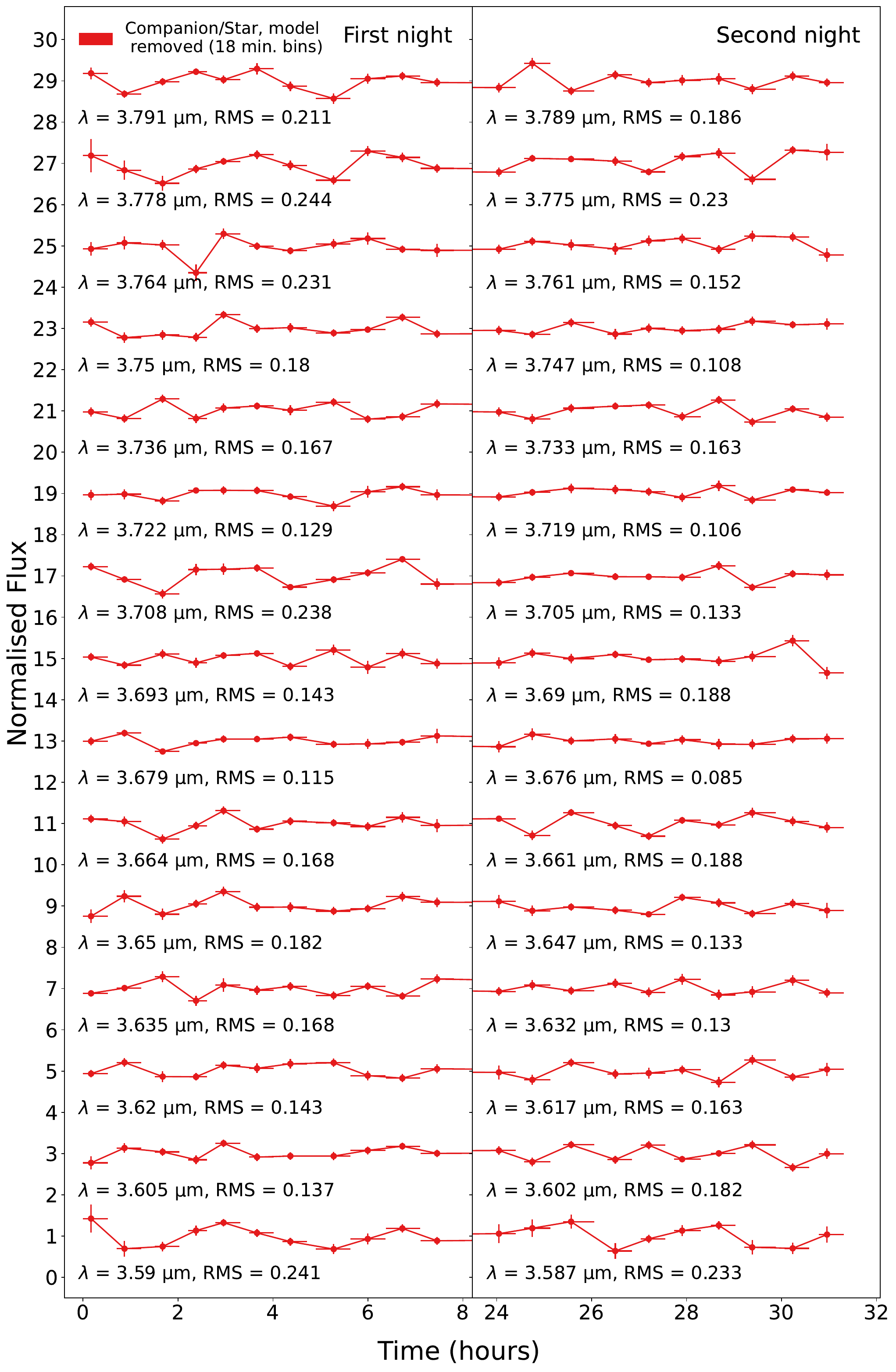}}}%
  \textcolor{white}{\frame{\includegraphics[width=0.5\textwidth]{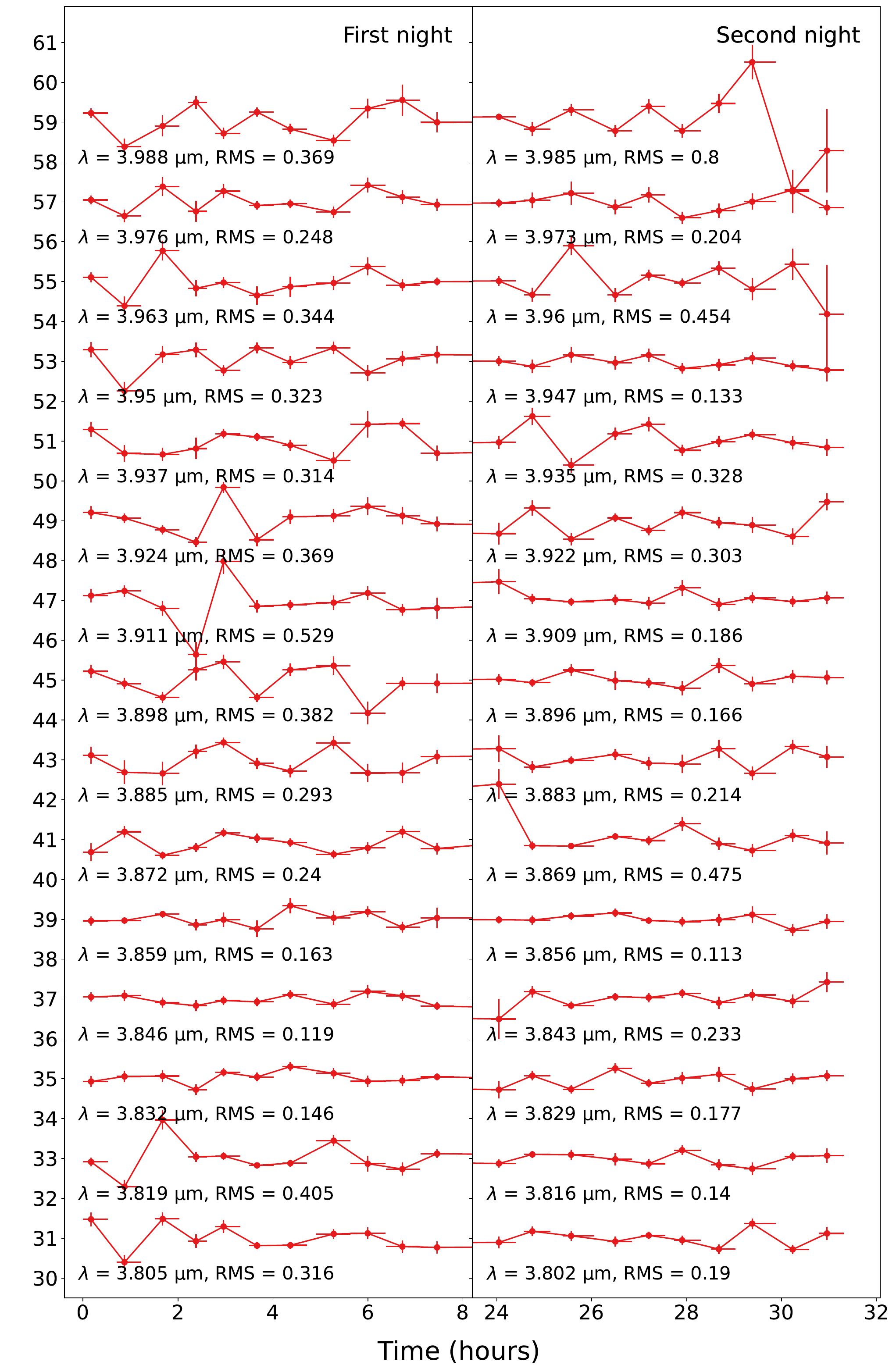}}}%
  \caption{The detrended differential light curves in each of the 30 individual wavelength channels that were combined to obtain white-light flux measurements for HD~1160~A and B. All light curves are binned to 18 minutes of integration time per bin, and each light curve is offset by 2 on the $y$-axis from the previous wavelength to spatially separate them in the figure.}
  \label{fig:p3_ch4_lc_30_both_nights}
\end{figure*}

We used a multiple linear regression \citep[as implemented in the scikit-learn Python package,][]{scikit-learn} to simultaneously fit these six decorrelation parameters to the differential white-light curve of HD~1160~B. This process was carried out for the light curve on each night separately, in case the systematics induce different trends on each night. The resulting model fits are shown in dark green in the top panels of Figure~\ref{fig:p3_ch4_detrended_lc_both}, overplotted on the raw differential white-light curves (in grey). The corresponding coefficients and intercept of these two models are given in Table~\ref{table:p3_ch4_lin_reg_coeffs}. We detrended the two differential white-light curves by dividing by these linear regression models, respectively. The final detrended differential white-light curves are plotted in the bottom panels of Figure~\ref{fig:p3_ch4_detrended_lc_both}, alongside the raw differential white-light curves for comparison (in light purple, reproduced from the bottom panel of Figure~\ref{fig:p3_ch4_raw_lc_both}). \citet{2023MNRAS.520.4235S} also presented detrended differential light curves for the first night in each of the 30 individual wavelength channels that were combined to obtain white-light flux measurements for HD 1160 A and B. To allow a comparison to their results, we also produced the detrended differential light curves in each wavelength channel for each night. These light curves are shown in Figure~\ref{fig:p3_ch4_lc_30_both_nights}, binned to 18 minutes of integration time per bin. The small differences in the wavelengths of each channel between nights arise from the different wavelength solution required for the wavelength calibration of each night of data (see Section~\ref{p3_ch4_nales_processing}).

\begin{flushleft}
\begin{table}
\caption{The decorrelation parameters $x_{i}$ included in the linear regression used to detrend the differential white-light curve of HD~1160~B at each epoch. The resulting linear model fit was given by $y = (\sum_{i=1}^{n}c_{i}x_{i}) + c_{0}$, where $c_{0}$ is the intercept and $c_{i}$ are the coefficients of each parameter. The parameters are ordered by the magnitude of the corresponding coefficients on the first night.}
\begin{tabular}{p{0.3\columnwidth}p{0.27\columnwidth}p{0.29\columnwidth}}
\hline
Parameter ($x_{i}$)&Value, 1st night ($c_{i}$)&Value, 2nd night ($c_{i}$)\\
\hline
Airmass&0.34590456&-0.46758893\\
Air temperature&0.11689032&-0.12725877\\
Star x-position&0.04596314&-0.08979261\\
Star y-position&-0.04426898&0.02947633\\
Companion x-position&-0.02499303&0.00674422\\
Companion y-position&0.01966123&-0.01849457\\
\hline
Intercept ($c_{0}$)&-1.23206397&5.45860550\\
\hline
\end{tabular}
\label{table:p3_ch4_lin_reg_coeffs}
\end{table}
\end{flushleft}

\subsection{Period analysis and light curve precision}\label{p3_ch4_results_periodograms}
\citet{2023MNRAS.520.4235S} identified sinusoidal-like variability in the first night of the detrended differential white-light curve of HD~1160~B and produced a Lomb-Scargle periodogram to search for periodicity. They then fit a sinusoid to the light curve and measured the period of this variability as 3.239~hours. This trend is still present in the first night of our new light curve (Figure~\ref{fig:p3_ch4_detrended_lc_both}). However, while some individual data points appear to deviate from a flat line, it is not visually clear whether or not the detrended differential white-light curve from the second night is also variable. The maximum normalised flux is 1.07, but the RMS of the light curve is 0.035. We therefore carried out a similar analysis to \citet{2023MNRAS.520.4235S} to search for periodic variability using periodograms.

\begin{figure*}
	\includegraphics[width=\textwidth]{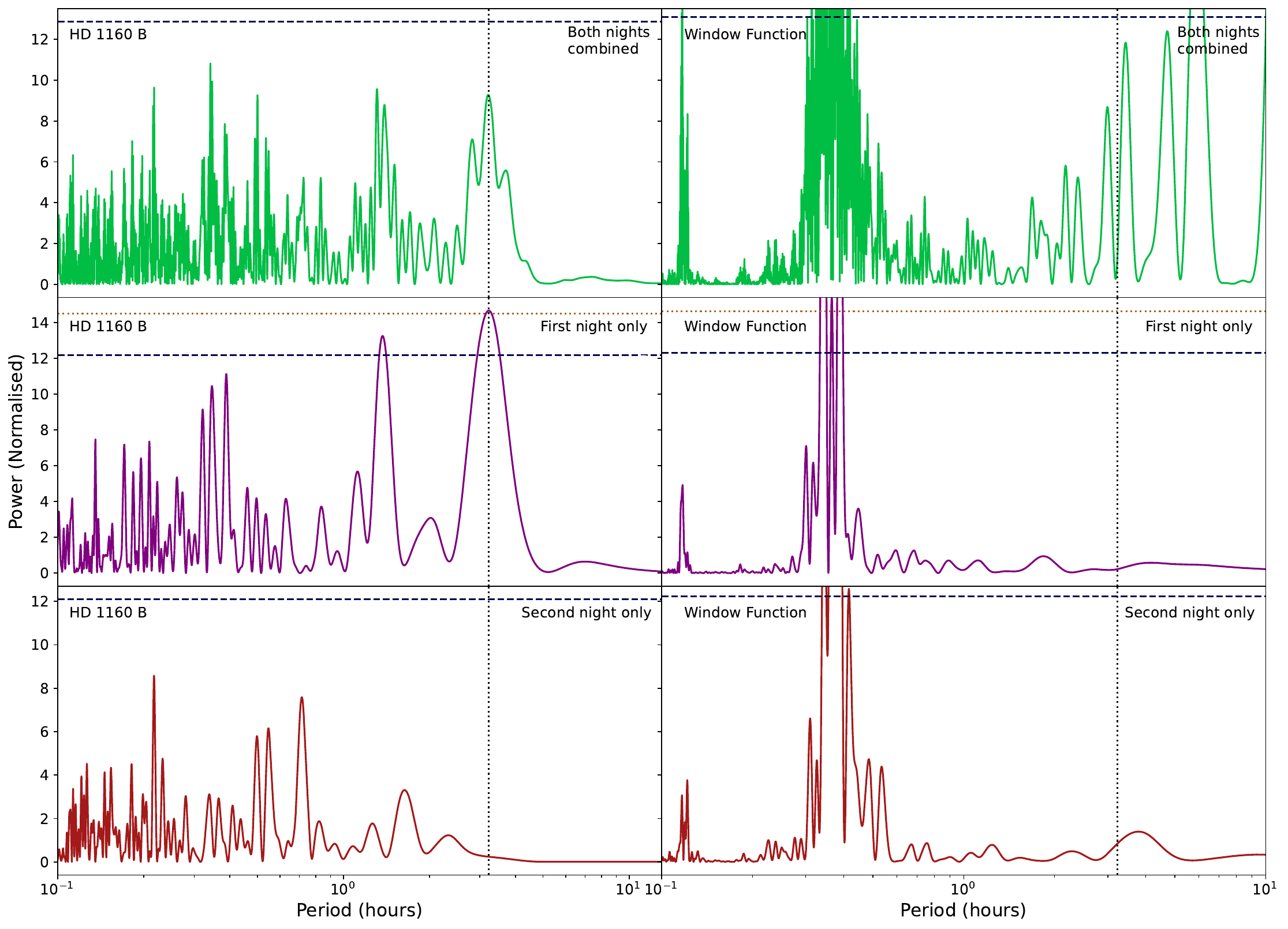}
    \caption{The left-hand panels show the Lomb-Scargle periodograms produced using the unbinned, detrended differential white-light curves from Figure~\ref{fig:p3_ch4_detrended_lc_both}. The top panel shows the periodogram for the full light curve over both nights, whereas the centre and bottom panels are those for the light curves of the first and second nights only, respectively. The right-hand panels show the periodograms of the corresponding window functions. The blue dashed lines indicate the power threshold corresponding to a false-alarm probability of 0.1 (10\%), and the horizontal brown dotted line is that for a false-alarm probability of 0.01 (1\%). The vertical dotted line highlights the 3.227~h period of the strongest peak in the periodogram for the first night.}
    \label{fig:p3_ch4_lomb_scargle}
\end{figure*}

We produced periodograms for the unbinned detrended differential white-light curve using the Lomb-Scargle algorithm \citep{1976Ap&SS..39..447L, 1982ApJ...263..835S}. These are shown in the left-hand panels of Figure~\ref{fig:p3_ch4_lomb_scargle}; the top panel was produced using both nights combined, while the centre and bottom panels show the periodograms produced using only the data from the first and second nights, respectively. Each of these power spectra has been normalised by dividing them by the variance of the data points in the light curve \citep{1986ApJ...302..757H}. The blue dashed line is the power threshold that corresponds to a false-alarm probability of 0.1 (i.e. 10\%), and the horizontal brown dotted line on the periodogram for the first night is the power threshold for a false-alarm probability of 0.01 (i.e. 1\%). 

We find that the strongest peak in the periodogram of the first night is at approximately the same period as \citet{2023MNRAS.520.4235S}, with a period of 3.227~hours, peak power of 14.67, and false-alarm probability of 0.009. This slight difference in period is due to the different normalisation used here and the different linear regression model produced by not including wind speed and wind direction as decorrelation parameters. The second strongest peak in this periodogram, with a period of 1.370~hours, peak power of 13.26, and false alarm probability of 0.035, does not appear to be harmonic with the strongest peak. However, there are no significant peaks present in the periodograms of the second night or of both nights combined. All of the features in these periodograms have false-alarm probabilities greater than 0.5. When the light curves are combined, the periodicity in the first night appears to be diluted by an absence of constructive addition from periodicity in the second night, causing there to be no significant peaks in the combined periodogram.

As the data are irregularly sampled, with large gaps due to the nodding pattern and the break between the two nights, we also produced periodograms for the window functions for each night and both nights together to check for potential artefacts arising from this irregular sampling \citep[e.g.][]{2019A&A...623A..24F, 2021ApJ...906...64A}. The window functions were calculated by producing an evenly-sampled array consisting of ones at times where data exists and zeros where it does not. The periodograms of these window functions are shown in the right-hand panels of Figure~\ref{fig:p3_ch4_lomb_scargle}. We find that there are no significant peaks at periods >1~hr in the window function periodograms for the first and second nights individually. This suggests that the strong peak that we detect at 3.227~hours in the periodogram of the first night light curve is not an artefact caused by the irregular sampling of the data. In both cases, there are significant peaks present at shorter periods (<1 hour), which is likely a reflection of the nodding pattern used when obtaining the data. These are also present for the both nights combined case along with several peaks at longer periods, which we interpret as harmonics of the nodding that appear due to the large gap between the two nights, but none align with the 3.227~hour peak from the first night.

\begin{figure}
	\includegraphics[width=\columnwidth]{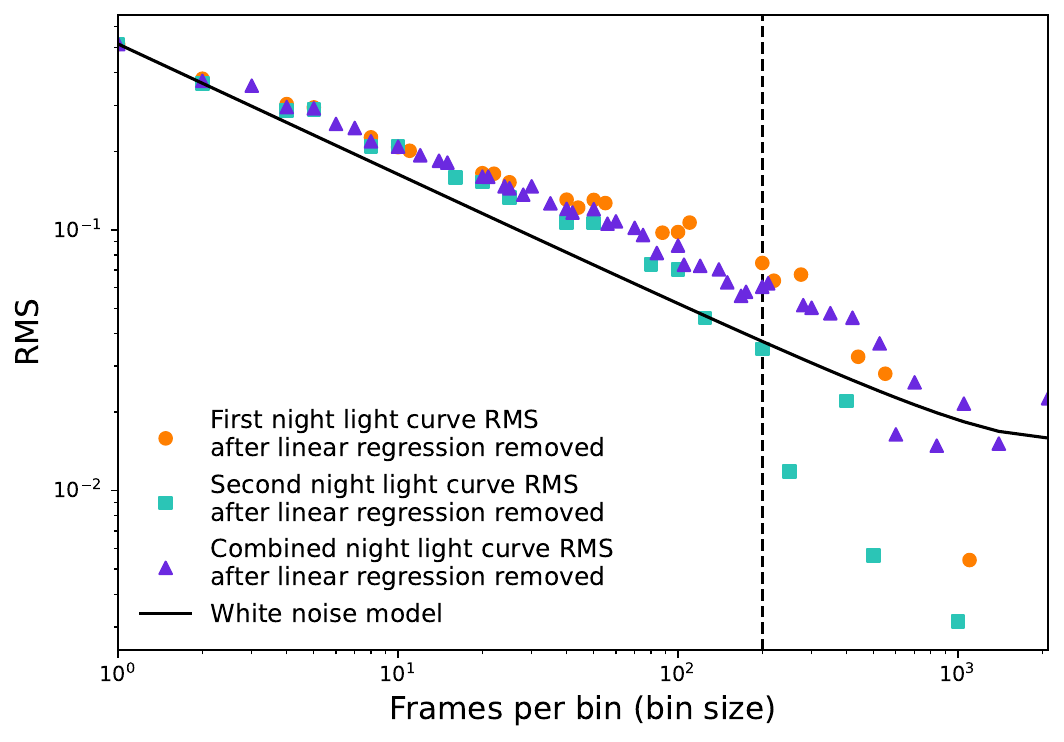}
    \caption{The RMS of the binned detrended differential white-light curve of HD~1160~B for the first and second nights, respectively, without the removal of any periodic variability. The theoretical white noise model as a function of bin size is also shown. This was calculated using the bin sizes used for the both nights combined light curve. The vertical dashed line indicates a bin size of 200, as used for our binned white-light curves in Figures~\ref{fig:p3_ch4_raw_lc_both}~and~\ref{fig:p3_ch4_detrended_lc_both}.}
    \label{fig:p3_ch4_rms_plot_no_rem}
\end{figure}

We also carried out a comparative analysis of the precision achieved in each detrended differential white-light curve. When estimating the precision achieved for the first night, \citet{2023MNRAS.520.4235S} first fitted and removed the observed periodic variability signal from the light curve. They did this using a non-linear least squares approach, assuming that it followed a sinusoidal trend and using the period of the highest peak in the periodogram as an initial guess for the fit. They then measured the precision using the residual light curve. As we do not detect a clear periodicity in the light curve from the second night, we could not do this here if we wished to compare the precision achieved on each night. We therefore instead performed our assessment of the precision using the detrended differential white-light curves from each night, and both nights combined, noting that any variations intrinsic to HD~1160~B would make these values appear higher and therefore above the true limiting precision. We did this by following the approach used by \citet{2011ApJ...733...36K} and \citet{2023MNRAS.520.4235S} for assessing the impact of correlated noise on time-series data. First, we binned our detrended differential white-light curves into a range of different bin sizes. We then renormalised the resulting binned light curves and subtracted one to centre them around zero, before measuring the root mean square (RMS) of each one. We plot these values as a function of bin size for each light curve in Figure~\ref{fig:p3_ch4_rms_plot_no_rem}. The black line shows the expectation of independent random numbers with increasing bin size, $\sigma_{N}=\sigma_{1}N^{-0.5}[M/(M-1)]^{0.5}$, where $N$ is the bin size and $M$ is the number of bins \citep{2011ApJ...733...36K}. If we take the RMS values at each night for the bin size that we used for our binned white-light curves in Figures~\ref{fig:p3_ch4_raw_lc_both}~and~\ref{fig:p3_ch4_detrended_lc_both} (i.e. 200 frames per bin, or 18 minutes of integration time), we find RMS values of 0.075 and 0.035 for the first and second nights, respectively. The RMS value at this bin size for the light curve covering both nights combined is 0.060. The higher RMS for the first night (and both nights combined) reflects the higher variability that we see here compared to the second night. We discuss these results further in Section~\ref{p3_ch4_variability_discussion}.

%% file: sections/06_spectral_analysis.tex
\section{Spectral analysis of HD~1160~B}\label{p3_ch4_spectral_results}
In addition to investigating the brightness fluctuations of HD~1160~B, we also extracted its spectrum on each night to allow us to characterize its physical properties through the fitting of atmospheric models.
\subsection{Spectral extraction}\label{p3_ch4_spectral_extraction_fluxcalib}
We measured the contrast between host star HD~1160~A and companion HD~1160~B in each wavelength channel using the aperture photometry of each object obtained in Section \ref{p3_ch4_section:aper_phot}. To do this, we took the median combination of these flux measurements over the time sequence, producing single flux measurements for the companion and the star at each wavelength. As with the time-dependent fluxes obtained in Section \ref{p3_ch4_4_diff_lcs_binning}, the errors on each measurement were calculated as the 1.48 $\times$ MAD of the fluxes in each wavelength channel (bin), divided by the square root of the number of frames per channel minus one. Next, we divided the companion flux at each wavelength by that of the star to produce a contrast spectrum. We carried out this process separately for each night of data.

\begin{figure}
	\includegraphics[width=\columnwidth]{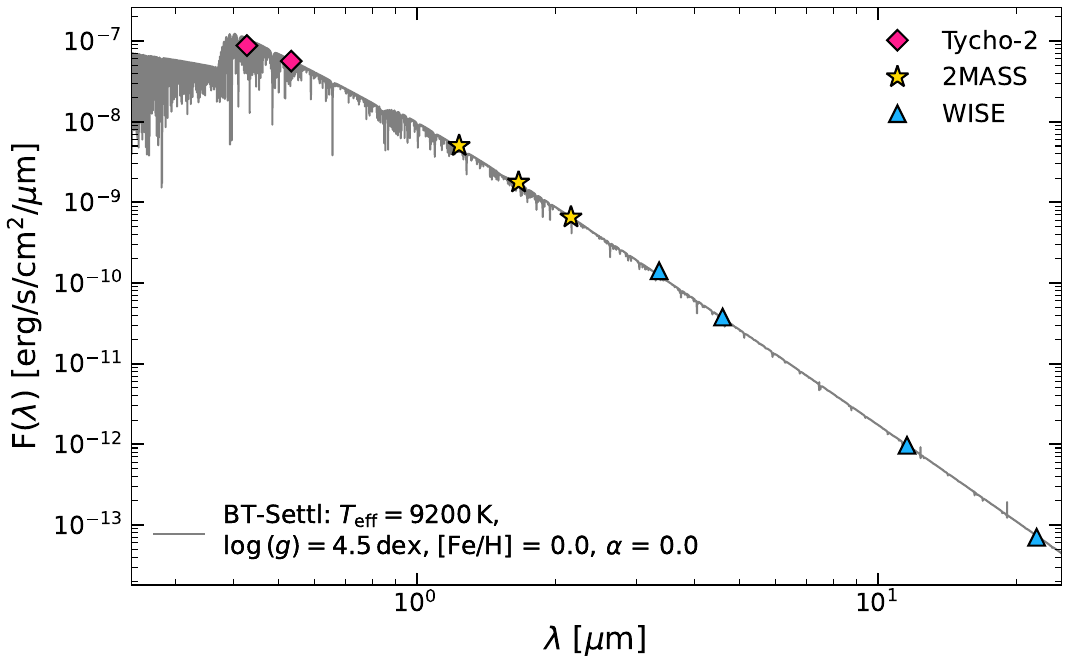}
    \caption{Literature photometry of the host star HD~1160~A from the Tycho, 2MASS, and WISE catalogues. The grey line shows the model fit to this photometry. The model has been convolved to a resolution of R~=~100,000 for visual purposes. The uncertainties are shown but are much smaller than the symbols.}
    \label{fig:p3_ch4_stellar_model_plot}
\end{figure}

\begin{figure*}
	\includegraphics[width=\textwidth]{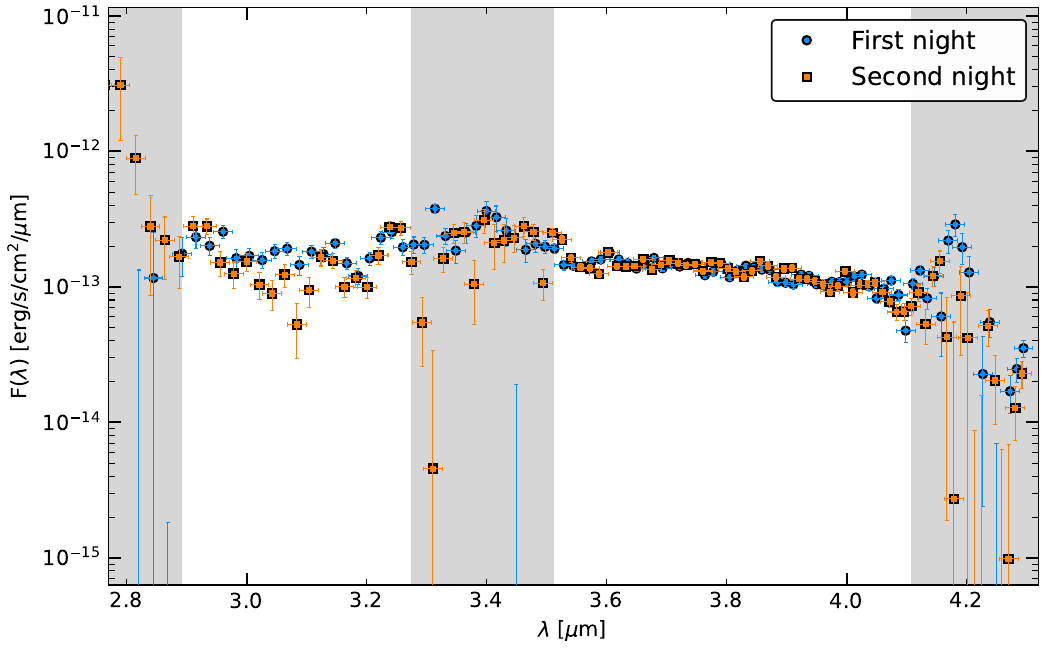}
    \caption{The flux calibrated spectrum of HD~1160~B obtained with LBT/ALES. Data points from the first and second nights are shown in blue and orange, respectively. The shaded regions indicate data points which are not suitable for analysis due to contamination arising from overlapping spectral traces or absorption caused by the carbon-carbon bonds in the glue layer of the dgvAPP360. The wavelength channels used for the variability analysis are those in the 3.59-3.99~$\upmu$m wavelength range.}
    \label{fig:p3_ch4_sed_plot}
\end{figure*}

We then converted our contrast spectra of HD~1160~B on each night into physical flux units by multiplying them by a flux calibrated spectrum of HD~1160~A. To do this, we used the Virtual Observatory SED Analyzer \citep[VOSA,][]{2008A&A...492..277B} to plot the Spectral Energy Distribution (SED) of HD~1160~A, including literature data from the 2MASS (\citealt{2006AJ....131.1163S}), Tycho-2 \citep{2000A&A...355L..27H, 2000A&A...357..367H}, and WISE (\citealt{2010AJ....140.1868W}) catalogues. We assumed a distance of 120.7~pc and an extinction of A\textsubscript{V} = 0.16 mag \citep{2016A&A...595A...1G, 2023A&A...674A...1G}. The SED was dereddened using the extinction law of \citet{1999PASP..111...63F} and \citet{2005ApJ...619..931I}. Using a $\chi^2$ test to fit the SED with a grid of BT-Settl models \citep{2011ASPC..448...91A, 2012RSPTA.370.2765A}, we selected a model with effective temperature T\textsubscript{eff}~=~9200~K, surface gravity log(g)~=~4.5~dex, metallicity [Fe/H] =~0.0, and alpha element abundance $\alpha$~=~0.0, consistent with that found by \citet{2020MNRAS.495.4279M} using the same approach. The literature photometry of HD~1160~A and this model are shown in Figure \ref{fig:p3_ch4_stellar_model_plot}. We then convolved this model to the resolution of ALES \citep[R$\sim$40,][]{2018SPIE10702E..0CS} and evaluated it at the wavelengths of our observations, before multiplying it by our contrast measurements to produce a flux calibrated spectrum of HD~1160~B on each night. These spectra are shown in Figure~\ref{fig:p3_ch4_sed_plot}, with the flux measurements from the first and second nights in blue and orange, respectively. The shaded areas indicate regions in the observed 2.8-4.2~$\upmu$m wavelength range where the data is unreliable and excluded from our analysis, due to contamination from the neighbouring spaxels or the dgvAPP360 glue absorption.

We note that while the spectra from each night are in good agreement in the wavelength region redward of the dgvAPP360 glue absorption, there appears to be a slight offset between the two nights at 3-3.2~$\upmu$m, which we discuss further in Section \ref{p3_ch4_spectral_discussion}.

\subsection{Spectral fitting}\label{p3_ch4_spectral_fitting}
Once we had obtained a flux calibrated spectrum of HD~1160~B for each night, we fit this data with atmospheric models to characterize its physical properties. We used a set of BT-Settl grid models \citep{2011ASPC..448...91A, 2012RSPTA.370.2765A, 2013MSAIS..24..128A} which were downloaded from the Spanish Virtual Observatory (SVO) Theory Server\footnote{\url{http://svo2.cab.inta-csic.es/theory/newov2/}}. We restricted the models to those with effective temperatures between 400~K and 4600~K, surface gravities between 3.5 and 5.0~dex, metallicities between $-0.5$ and 0.5, and an $\alpha$-enhancement of 0. The grid step sizes for temperature and surface gravity were 100~K and 0.5~dex, respectively, and metallicity could have values $-0.5$, 0, 0.3, or 0.5. We chose to restrict the range of possible surface gravities to these values based on the predicted physical limitations of objects with ages and masses within the ranges found for HD~1160~B in the recent literature, which are 10-125~Myr and $\sim$20~M\textsubscript{Jup} to 123~M\textsubscript{Jup}, respectively \citep{2017ApJ...834..162G, 2019AJ....158...77C, 2020MNRAS.495.4279M}. According to the isochrones and evolutionary tracks of the BT-Settl models, the surface gravities of objects with ages and masses within these constraints should always be $\geq$3.5 and $\leq$5.0 \citep{2015A&A...577A..42B, 2016ApJ...829...39S}. \citet{2016A&A...587A..56M} did previously use a higher age upper limit of 300~Myr for the HD~1160 system, which would allow a HD~1160~B surface gravity of up to log(g)~=~$\sim$5.2, but \citet{2017ApJ...834..162G} later found that such high ages were not consistent with the properties of the host star.

\begin{figure*}
	\includegraphics[scale=0.95]{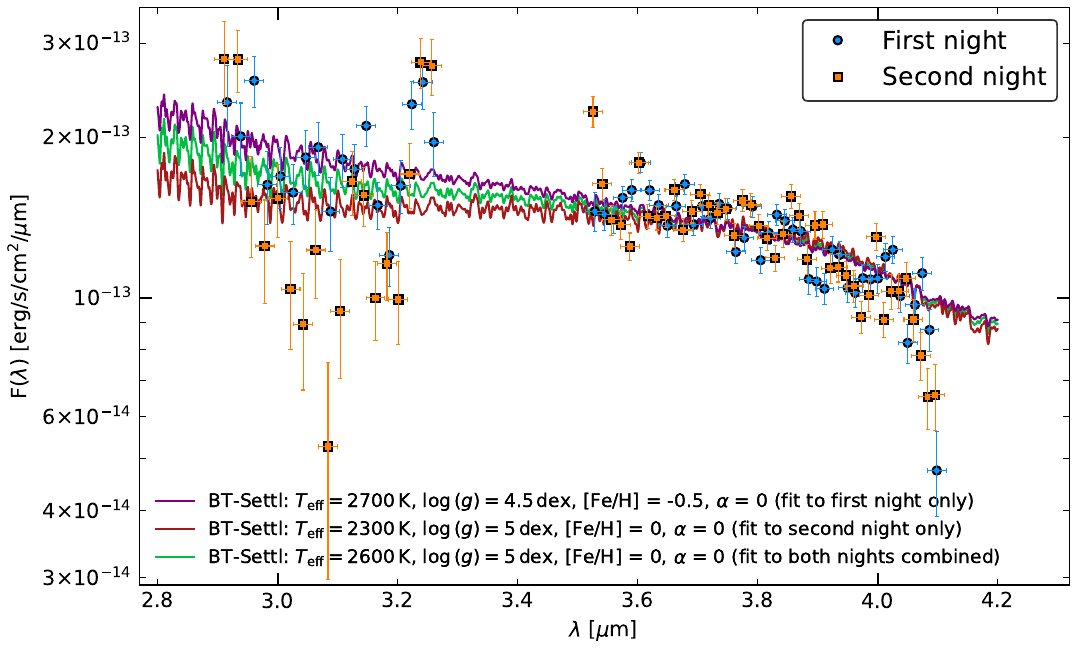}
    \caption{The best-fitting models to the flux-calibrated ALES spectrum of HD~1160~B. The large difference in the temperatures of the best-fitting models appears to arise from the difference in flux between the two nights in the 3.0-3.2 $\upmu$m region. Data points in the shaded regions in Figure~\ref{fig:p3_ch4_sed_plot} were not included in these fits and are therefore not shown.}
    \label{fig:p3_ch4_model_fit_plot}
\end{figure*}

Each model was convolved to the R$\sim$40 spectral resolution of ALES and sampled at the wavelengths of our spectral data points. By fitting each model to the data, we then determined the scaling factor that minimises the Euclidean norm of the residual vector between the two i.e. the value multiplied by each model to best match it to the companion spectrum, and calculated the $\chi^2$ value for each fit accounting for the errors on each data point \citep[e.g.][]{2020MNRAS.492..431B, 2021MNRAS.506.3224S}. When calculating the $\chi^2$ values of model fits to high-contrast IFS data, it is important to consider the effects of spectral covariance arising from oversampling during the spectral extraction process and the wavelength-dependent behaviour of speckle noise \citep{2016ApJ...833..134G}. We accounted for spectral covariances in the ALES IFS data by following the method described by \citet{2016ApJ...833..134G} to produce spectral covariance matrices for each night of data and apply them in our $\chi^2$ calculation. The model that produced the smallest $\chi^2$ value was then taken as the best-fitting model to the data. When performing this fitting procedure we excluded the data points in the shaded regions of Figure~\ref{fig:p3_ch4_sed_plot}, which were not suitable for analysis as described in Section \ref{p3_ch4_spectral_extraction_fluxcalib}. We performed the fitting process three times; once each for the spectra from the first and second nights separately, and a third time considering both nights of data together. The best-fitting model in each case is shown overplotted on the companion spectrum in Figure \ref{fig:p3_ch4_model_fit_plot}. The best-fitting model to the first night of data alone has T\textsubscript{eff} = 2700~K, log(g) = 4.5~dex, and metallicity [Fe/H] = $-0.5$ (purple line, Figure \ref{fig:p3_ch4_model_fit_plot}). When the second night of data is considered alone, the best-fitting model instead has solar metallicity and is slightly cooler, with T\textsubscript{eff} = 2300~K, log(g) = 5.0~dex (red line). This is likely due to the lower flux recorded in the 3-3.2~$\upmu$m region of the spectrum on this night. The effective temperature of the best-fitting model to both nights of data then lies between the two, as would be expected, with T\textsubscript{eff} = 2600~K, log(g) = 5.0~dex, and solar metallicity (green line). Using the $\chi^2$ values of each model fit as weights, we also calculated the weighted means and sided variance estimates (i.e. statistical errors) of these atmospheric parameters using the approach of \citet{2010ApJ...710.1142B, 2010ApJ...725.1405B, 2016ApJ...829...39S}. These results are presented in Table~\ref{table:p3_ch4_model_physical_parameters}. The weighted means and their uncertainties are biased in some cases, where the preferred model fits lie at the edge of the allowed parameter range. We therefore instead report upper/lower limits in these instances.

We further inferred estimates of the radius and luminosity of the companion using the scaling factor for each model, which is equal to the squared ratio of the companion's radius and distance. As the distance to the HD~1160 system is well-known \citep[120.7$\pm$0.5~pc][]{2016A&A...595A...1G, 2023A&A...674A...1G}, we are able to solve for radius. The luminosity can then be inferred by integrating each model over its full wavelength range and multiplying by $4\pi$ times the radius squared. These values are also shown in Table~\ref{table:p3_ch4_model_physical_parameters}, where the reported uncertainties are again the statistical errors. These results are discussed further in Section \ref{p3_ch4_spectral_discussion}.

\begin{flushleft}
\begin{table}
\caption{The physical properties of HD~1160~B as derived by fitting BT-Settl models to its spectrum from the first night, the second night, and both nights combined. These values are the weighted means calculated based on the $\chi^2$ values of each model fit. The uncertainties reported here are only the statistical errors based on sided variance estimates. Where the fitting procedure tends to prefer models at the edge of the allowed parameter range, we instead report upper/lower limits. The bottom part of the table shows the estimated mass ranges for HD~1160~B and the corresponding mass ratios $q$ relative to HD~1160~A, as found in Section~\ref{p3_ch4_companion_mass} by evaluating BT-Settl isochrones at our luminosity values. These ranges are wide due to the wide age range considered, 10-125~Myr.}
\begin{tabular}
{p{0.26\columnwidth}p{0.18\columnwidth}p{0.18\columnwidth}p{0.18\columnwidth}}
\hline
Property&First night&Second night&Both nights\\
\hline
T\textsubscript{eff} (K)&2794{\raisebox{0.5ex}{\tiny$\substack{+115 \\ -133}$}}&2279{\raisebox{0.5ex}{\tiny$\substack{+79 \\ -157}$}}&2554{\raisebox{0.5ex}{\tiny$\substack{+49 \\ -93}$}}\\

log(g) (dex)&$\geq$4.08&$\geq$4.41&$\geq$4.49\\

Metallicity&$\leq$0.27&0.00{\raisebox{0.5ex}{\tiny$\substack{+0.41 \\ -0.00}$}}&$\leq$0.05\\

Radius (R\textsubscript{Jup})&1.46{\raisebox{0.5ex}{\tiny$\substack{+0.08 \\ -0.06}$}}&1.77{\raisebox{0.5ex}{\tiny$\substack{+0.12 \\ -0.05}$}}&1.59{\raisebox{0.5ex}{\tiny$\substack{+0.06 \\ -0.03}$}}\\

log(L/L$_\odot$)&-2.91{\raisebox{0.5ex}{\tiny$\substack{+0.03 \\ -0.03}$}}&-3.09{\raisebox{0.5ex}{\tiny$\substack{+0.03 \\ -0.06}$}}&-2.99{\raisebox{0.5ex}{\tiny$\substack{+0.02 \\ -0.03}$}}\\
\hline
Mass (M\textsubscript{Jup})&18.0-81.4&15.5-65.4&17.1-72.1\\
Mass ratio $q$&0.008-0.038&0.007-0.030&0.008-0.034\\
\hline

\end{tabular}
\label{table:p3_ch4_model_physical_parameters}
\end{table}
\end{flushleft}

%% file: sections/07_pepsi.tex
\section{Characterizing HD~1160~A with PEPSI}\label{p3_ch4_PEPSI_method_results}
\begin{figure*}
	\includegraphics[width=\textwidth]{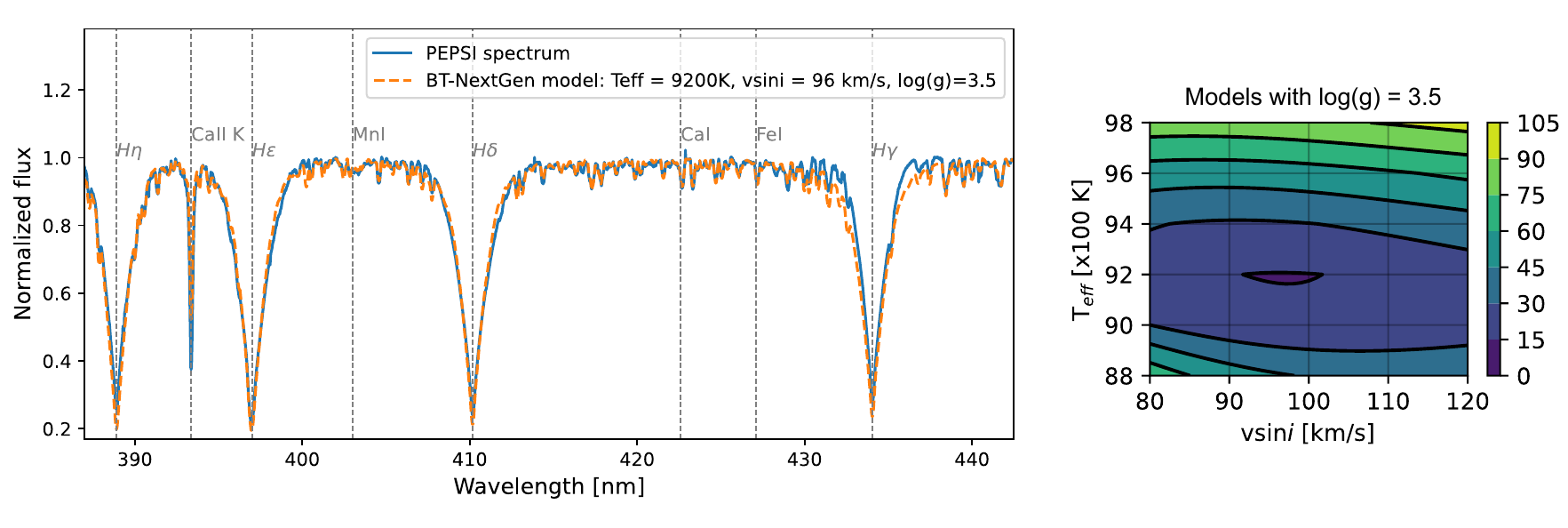}
    \caption{The left panel shows the PEPSI spectrum of the host star HD~1160~A in blue, overplotted with the best-fitting model from BT-Nextgen in orange. The fitting process was carried out for the region of the spectrum covering 392-429~nm. The contour plot in the right panel shows the $\chi^2$ distribution for several temperatures and $\varv \sin i$ at fixed log(g) of 3.5.}
    \label{fig:p3_ch4_pepsi_spec}
\end{figure*}

In addition to characterizing HD~1160~B using the data obtained with ALES+dgvAPP360, the simultaneous high resolution PEPSI spectrum of the HD~1160 system in the optical (383-542\,nm) further allows us to assess the properties of the host star HD~1160~A, which was originally classified as an A0\,V star by \cite{1999mctd.book.....H}. Although HD~1160~C lies at an angular separation far beyond the 2.25$\arcsec$ diameter of the PEPSI fiber, HD~1160~B lies within this fiber diameter at a separation or $\sim$0.78$\arcsec$, so the obtained PEPSI spectrum contains the spectra of both HD~1160~A and B. However, the contrast between the two is very large: $7.72\pm0.01$ mag in the 1.25~$\upmu$m J-band, and even larger at the shorter wavelengths covered by PEPSI \citep{2017ApJ...834..162G}. We therefore assumed that the contribution of HD~1160~B to the PEPSI spectrum was negligible and treated the PEPSI spectrum as solely that of HD~1160~A (see Figure~\ref{fig:p3_ch4_pepsi_spec}).

To estimate the properties of HD~1160~A, we compared the spectrum to BT-NextGen atmospheric models, which are computed with the use of the PHOENIX code \citep{1999ApJ...512..377H, 2012RSPTA.370.2765A}. The input parameters for the model spectra were effective temperature (T$_\text{{eff}}$), surface gravity (log(g)), and metallicity, the latter of which was taken as solar for HD~1160~A. The models were convolved to the resolution of the PEPSI instrument and broadened by the rotation of the star ($\varv \sin i$). We identified the best fit values for these parameters by determining the $\chi^2$ values for a grid of models, varying T$_\text{{eff}}$ (8800-9800\,K in steps of 200\,K), log(g) (1.5-4.5 in steps of 0.5), and $\varv \sin i$ (80-120\,km~s$^{-1}$ in steps of 1\,km~s$^{-1}$). The ranges of these parameters were chosen based on an initial visual inspection of the PEPSI spectrum using the digital spectral classification atlas of \citet{gray2000digital}. The model grid spectra were normalised with splines fitted at similar (continuum) points for a given T$_\text{{eff}}$. The same continuum points are used for a re-normalisation of the PEPSI spectrum with a spline to match the normalisation of the grid spectra. However, the shape of the Balmer lines appears to be inconsistent between lines, which is hard to explain with any intrinsic properties for this type of star \citep{gray2000digital}. We interpret this as a systematic error arising from residual fringing, and therefore excluded the region around the H$\beta$ and H$\gamma$ lines from the fitting procedure. The region of the spectrum that we used for the fitting process was therefore 392-429~nm.

The errors on the PEPSI data points given by the automated pipeline are on average 0.0003\% of the flux, which corresponds to an extremely high S/N of $\sim$\,330000 that we interpret as implausible since the PEPSI exposure time calculator requires $\sim$\,1.4\,years of exposure time to achieve this, while our exposure times were $\sim$\,14800\,s. Furthermore, the spectrum seems to contain a low level sinusoidal-like structure, which most likely arises from systematics introduced by the original normalisation performed by the automated pipeline. We therefore recalculated the error on each data point using the S/N instead measured from the normalised spectrum by taking the inverse of the standard deviation of the flux in the continuum, which gives S/N~=~$\sim$500 (or $\sim$0.2\% of the normalized flux). This value is then weighted by $\sqrt{F_{\text{i}}}$, where $F_\text{i}$ is the flux for a given wavelength point i, to calculate the observed errors for each wavelength point. 

We found that the resulting best fit model, taken as that with the lowest $\chi^2$ value (=~14.46), has T\textsubscript{eff}~=~9200$^{\,+\,200}_{\,-\,100}$\,K, $\varv \sin i$~=~96$^{\,+\,6}_{\,-\,4}$\,km~s$^{-1}$, and log(g)~=~3.5$^{\,+\,0.5}_{\,-\,0.3}$~dex, where the errors on these values are based on the model grid spacing and distribution of $\chi^2$ values. This corresponds to an A1\,IV-V classification for HD 1160\,A \citep{2000asqu.book.....C}. This best-fitting model is shown overplotted on the PEPSI spectrum of HD~1160~A in the left panel of Figure~\ref{fig:p3_ch4_pepsi_spec}. The right panel then shows the $\chi^2$ distribution for models with a log(g) of 3.5 over temperature and $\varv \sin i$. The relatively high $\chi^2$ values, even for the best fit model, are due to normalisation differences between the model and the spectrum, the very small errors on the flux, and the large grid separation for T$_\text{{eff}}$ and log(g) for BT-Nextgen models. We discuss these results further in Section~\ref{p3_ch4_discussion_pepsi}.

%% file: sections/08_discussion.tex
\section{Discussion}\label{p3_ch4_discussion}
\subsection{HD~1160~B light curves}\label{p3_ch4_variability_discussion}
\subsubsection{The variability of HD~1160~B}\label{p3_ch4_variability_interp_discussion}
In Section~\ref{p3_ch4_results_periodograms}, we recovered the high-amplitude $\sim$3.2~h periodic variability signal identified by \citet{2023MNRAS.520.4235S} in the first night of the detrended differential white-light curve of HD~1160~B. We also found that some data points in the light curve from our additional night deviate from equilibrium flux, albeit with a smaller amplitude. However, we do not identify any periodic signals in the light curve from this second night, nor in the full light curve covering both epochs. In both of these cases, all peaks in their respective periodograms lie well below the 1\% level.

Let us first consider the case that HD~1160~B is variable. There are several physical mechanisms that could potentially explain the decrease or absence of variability that we see on the second night. Variability in substellar objects arises from clouds or other atmospheric features, such as magnetic spots if the object is of higher mass, rotating in and out of view over their rotation periods \citep[e.g.][]{2001ApJ...556..872A, 2010ApJ...718..502M, 2012MNRAS.427.3358G, 2014ApJ...793...75R, 2015ApJ...799..154M, 2019ApJ...874..111T, 2022ApJ...924...68V}. Where multiple such features with different sizes are present in the atmosphere of a companion at different locations, the resulting variability signal can appear irregular in amplitude, phase, and/or periodicity \citep[e.g.][]{2003AJ....126..348T, 2016ApJ...830..141L}. It is possible that we are seeing this effect in the full light curve of HD~1160~B; if its true rotation period is in fact longer than $\sim$3.2~h (and perhaps even longer than the baseline of a single epoch), then we could be viewing it at a different phase in its rotation on the second night. In this case, additional observations would be required to cover the full rotation period of HD~1160~B and verify whether or not these trends repeat. Regardless, a $\sim$3.2~h period is consistent with the fastest rotation periods of young isolated objects with a similar spectral type to HD~1160~B (i.e. late M- and early L-dwarfs), which have periods ranging from $\sim$2-72~h \citep[e.g.][]{2004A&A...419..703B, 2021ApJ...916...77P, 2022ApJ...924...68V}. Another possibility is that the difference in the level of variability is due to evolution in the surface features and atmospheric dynamics that cause the variability \citep[e.g.][]{2021MNRAS.502..678T}. Many studies have identified changing variability in the light curves of brown dwarfs and planetary-mass objects, including both long-term trends over hundreds of rotation periods and rapid light curve evolution from one night to the next or even between consecutive rotations \citep[e.g.][]{2002ApJ...577..433G, 2009ApJ...701.1534A, 2012ApJ...750..105R, 2013A&A...555L...5G, 2016ApJ...825...90K, 2017Sci...357..683A, 2021ApJ...906...64A, 2022AJ....164..239Z, 2024ApJ...965..182F}. If the rotation period of HD~1160~B is $\sim$3.2~h, it would have completed $\sim$5 rotations between the end of the observing sequence on the first night and the start of observations on the second night, which may be long enough for rapid evolution to have occurred. However, we also note that the first night covers 7.81~h which corresponds to only $\sim$2.5 rotations of HD~1160~B. It may be the case that this is insufficient to accurately derive the rotation period of HD~1160~B. Regardless, significant night-to-night changes are especially possible if the atmosphere has a banded structure with sinusoidal surface brightness induced by planetary-scale waves, as multiple bands with slightly different periods can give rise to a beating effect. For example, \citet{2017Sci...357..683A} found that an analytical model combining three sinusoids with different periods (corresponding to three atmospheric bands) produces a function that can rapidly fluctuate from low to high amplitudes. They found that this model matches the light curve evolution in their Hubble Space Telescope observations of an L/T transition brown dwarf, which shows low-amplitude variability on one day and high-amplitude variability on the next in a similar manner to HD~1160~B.

Several studies exploring the variability of substellar objects in different wavebands have further found that light curves can have similar shapes at different wavelengths, but with an offset in phase as different wavelengths probe different atmospheric pressures \citep[e.g.][]{2012ApJ...760L..31B, 2016ApJ...826....8Y, 2013ApJ...778L..10B, 2018AJ....155...95B, 2019AJ....157...89G, 2024ApJ...965...83M, 2024arXiv240304840P}. Since our differential light curve of HD~1160~B is a white-light curve integrated over a wide wavelength range, such wavelength-dependent phase offsets could lead to a `cancelling out' effect if they were of certain amplitudes. This effect could impact the light curve of HD~1160~B if its variability has different periods at different wavelengths, such that their phases mismatch at certain times.

\citet{2023MNRAS.520.4235S} highlighted that if HD~1160~B is a low-mass M-dwarf, a short-lived flaring period could be the cause of its $\sim$8.8\% semi-amplitude variability on the first night. If this is indeed the case for HD~1160~B, this would be consistent with both the high-amplitude variability seen on the first night and its decrease or absence on the second night. However, while flaring events of this magnitude have been observed in the infrared, they are expected to be rare \citep[e.g.][]{2012ApJ...748...58D, 2012MNRAS.427.3358G, 2012AJ....143...12T}.

We must now also consider the possibility that one or more unknown systematics could be responsible for the high-amplitude periodic variability that we see on the first night. However, it is not clear what systematic effect could induce such high-amplitude periodic variations on one night and not do so on the following night, given that the same methodology was applied to each epoch. The observing conditions were very similar and highly stable on both nights. If we consider the decorrelation parameters used in the detrending procedure (Figure~\ref{fig:p3_ch4_linreg_parameters}), we see that the airmass, companion position, and stellar position all follow approximately the same trends on each night. This would appear to rule out residual systematics arising from these parameters as the source of the light curve differences between each night. The air temperature does differ slightly in the second half of each night but is otherwise broadly similar, and any correlation arising from this difference is unlikely to be significant enough to explain what we see, particularly after the detrending process has been applied. One possibility could be that there is an additional systematic connected to the flexure of the ALES lenslet array as the telescope rotates. However, any such effects should already be accounted for by the inclusion of the pixel positions of HD~1160~A and B in the detrending process. An alternative parameter probing this flexure would be the pointing altitude of the telescope, which again follows the same trend on each night (and follows an inverse relationship to airmass), suggesting that this is unlikely to be the source of the differences between the two nights. Furthermore, \citet{2023MNRAS.520.4235S} found that the shape of the raw light curves is robust against issues arising from lenslet flexure by comparing data processed using wavelength calibration frames obtained at different pointing altitudes. We also note the sharp increase in flux in the raw differential light curve on the second night at $\sim$27~h. This feature appears to be connected to the large offset in the stellar (and companion) x-position at this time, and is accounted for in the linear regression and hence removed in the detrended light curve.

\begin{figure}
	\includegraphics[width=\columnwidth]{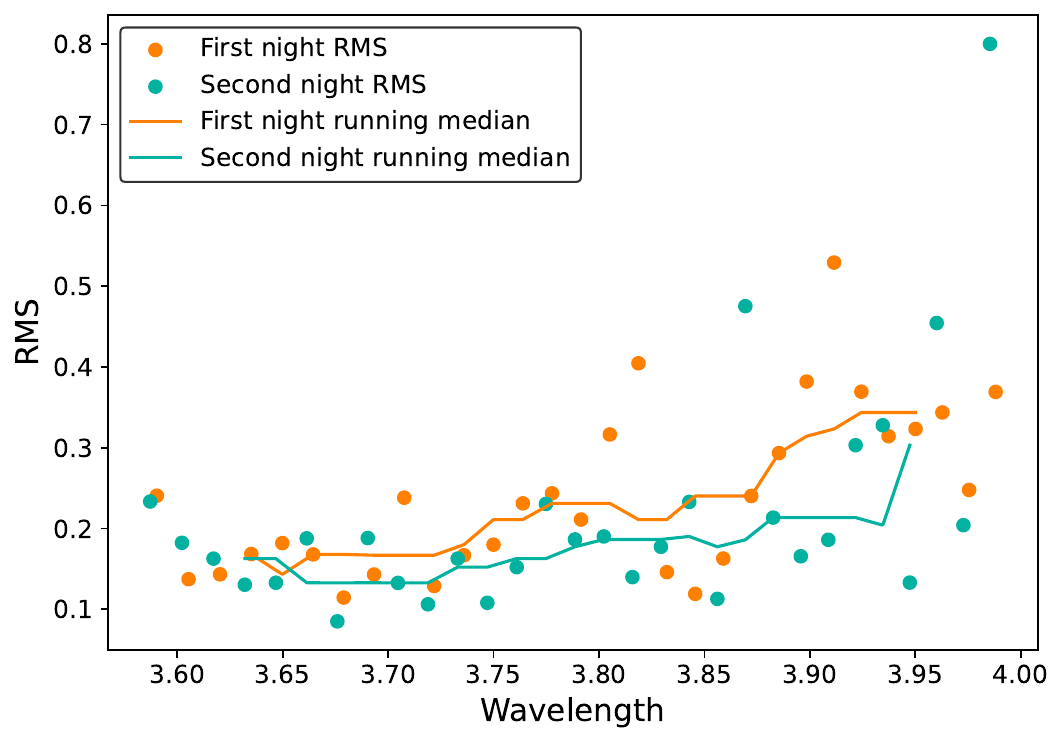}
    \caption{The RMS values, as a function of wavelength, for the 30 detrended differential light curves in the individual wavelength channels (shown in Figure~\ref{fig:p3_ch4_lc_30_both_nights}) that were combined to obtain the white-light curve. The binning is the same at 18 minutes of integration time per bin. The running medians of these values for each night are also shown, based on a window size of 7. The RMS increases with wavelength on both nights.}
    \label{fig:p3_ch4_rms_wavelength_plot}
\end{figure}

\citet{2023MNRAS.520.4235S} also used the RMS values of the detrended differential light curves in each of the 30 individual wavelength channels from the first night as a metric to search for wavelength-dependent trends. To allow a comparison to their results, we repeated this for the individual channel light curves on both nights (Figure~\ref{fig:p3_ch4_lc_30_both_nights}). These RMS measurements are shown as a function of wavelength in Figure~\ref{fig:p3_ch4_rms_wavelength_plot}, along with the running median on each night using a window size of 7. \citet{2023MNRAS.520.4235S} identified a tentative increase in the RMS towards longer wavelengths on the first night. We see this trend again on both the first and second nights, as indicated by the upward curves of the running median. Although this trend could potentially indicate an increase in variability with wavelength, this increase in RMS is more likely explained by noise due to the higher thermal background at longer wavelengths. There are outliers from this trend, which could in principle be explained by changes in the atmosphere of the companion between nights. However, the wavelengths at which these outliers occur are not consistent between nights and as the S/N in each individual wavelength channel is not high, we do not speculate further on their origin here.

Although the nature of the variability of HD~1160~B remains unclear, the additional night of variability monitoring presented here shows that this variability does not follow a simple periodic trend and highlights the complexities of interpreting the light curves of high-contrast substellar companions. Future ground-based observations will help to shed light on the trends in the light curves of HD~1160~B through additional epochs that provide a longer baseline when combined with our data, and the greater photometric precision provided by space-based facilities such as JWST could further help to constrain its variability amplitudes.

\subsubsection{Precision of vAPP differential spectrophotometric monitoring}\label{p3_ch4_diff_discuss_precision}
Our additional night of variability monitoring through differential spectrophotometric monitoring combined with the dgvAPP360 allows us to test whether this recently-developed technique can achieve the same precision at multiple epochs. \citet{2023MNRAS.520.4235S} found that this technique did not reach a systematic noise floor on the first night, suggesting that the precision would continue to improve with a longer baseline and increasing bin size. In Section~\ref{p3_ch4_results_periodograms}, we measured the RMS as a function of bin size for the detrended differential white-light curves on each night and both nights combined (Figure~\ref{fig:p3_ch4_rms_plot_no_rem}). As noted previously, the RMS trends for the first night and both nights combined cases do sit at a slightly higher level than on the second night, but this is expected as no astrophysical variability signal has been removed from either night and the variability is of higher amplitude on the first night. Aside from this offset, we see that the RMS follows the same trend on both nights; both decrease according to the trend of the white noise and do not plateau. This is also the case for the light curve covering both nights combined. This suggests that the data possesses similar noise properties at both epochs and therefore that the precision reached with this technique can be reliably repeated. The RMS of the detrended differential white-light curve from the second night is 0.035 in bins of 18~minutes, corresponding to a precision of 3.5\%. This is comparable to the 3.7\% precision measured by \citet{2023MNRAS.520.4235S} for the first night light curve with the same bin size, indicating that this precision level is repeatable over multiple epochs. Since we do not appear to reach the photon noise limit with these observations, future observations may be able to achieve a greater precision by improving the process used to detrend the differential light curve of the companion. This could be done by adapting more complex approaches used for exoplanet transmission spectroscopy studies, including techniques using Gaussian processes which do not assume that the systematics have any particular dependence on the telluric/instrumental parameters \citep[e.g.][]{2012MNRAS.419.2683G, 2013ApJ...772L..16E, 2018Natur.557..526N, 2020MNRAS.494.5449C, 2020AJ....160...27D, 2020AJ....160..188D, 2022MNRAS.510.3236P, 2023ARA&A..61..329A}. However, we note that robustly assessing the true deviation of these trends from the white noise model is difficult due to the possible astrophysical variability. Furthermore, although such techniques will help to minimise the risk of overfitting, transmission spectroscopy targets are typically pixel-stabilized. Thus, implementing them for non-stabilized differential spectrophotometry targets will require careful consideration.

\subsection{Spectral characterization of HD~1160~B}\label{p3_ch4_spectral_discussion}
In Section \ref{p3_ch4_spectral_extraction_fluxcalib}, we presented the extracted spectra of HD~1160~B for each night and noted an offset in the spectra of HD~1160~B from each night in the 3-3.2~$\upmu$m wavelength region, where the data points from the second night appear to lie slightly lower than those from the first night. The cause of this offset is unclear. The time-averaged flux measurements of the host star HD~1160~A on each night are consistent across all wavelengths, indicating that this is not caused by throughput differences arising from differences in the weather conditions between the two nights. A possibility is that this feature is astrophysical and arises from the intrinsic variability of HD~1160~B, with it appearing fainter in this wavelength region on the second night. In addition to this feature, the scatter of the datapoints at longer wavelengths (e.g. 3.6-4.0~$\upmu$m) is also larger than would be expected from the fitted models. If our uncertainties are correctly estimated, then this may also be due to the effects of variability; a greater scatter in the spectrum of HD~1160~B would be expected if its variability has different properties at different wavelengths. However, we cannot rule out that this scatter is a systematic effect.

Regardless of the origin of the offset at 3-3.2~$\upmu$m, it has a significant impact on the results of the atmospheric model fitting described in Section~\ref{p3_ch4_spectral_fitting}. This process produced significantly different values for the physical properties of HD~1160~B depending on whether the models were fit to the spectra from the first night alone, the second night alone, or both nights combined (Table~\ref{table:p3_ch4_model_physical_parameters}). The values for effective temperature T\textsubscript{eff} cover a particularly large range, from 2279{\raisebox{0.5ex}{\tiny$\substack{+79 \\ -157}$}}~K for the fit to the second night to 2794{\raisebox{0.5ex}{\tiny$\substack{+115 \\ -133}$}}~K on the first night (a >1$\sigma$ difference). The effective temperatures derived from the second night and both nights combined spectra are much cooler than those in the literature, but the higher temperature from the first night is consistent with previous measurements to within 1$\sigma$. \citet{2016A&A...587A..56M} determined a T\textsubscript{eff} for HD~1160~B of 3000$\pm$100~K through atmospheric modelling, consistent with the 3000-3100~K value found by \citet{2017ApJ...834..162G}, although the latter study noted that they could not rule out slightly cooler temperatures. Our value derived from the spectrum from the first night also overlaps with the 2800-2900~K range estimated by \citet{2020MNRAS.495.4279M}.

All three of our constraints for surface gravity log(g) are consistent with the 4.0-4.5 dex range estimated by \citet{2017ApJ...834..162G} (who also could not rule out slightly higher values), and consistent within 2$\sigma$ with \citet{2020MNRAS.495.4279M}, who estimated a lower log(g) of 3.5-4.0 dex. \citet{2016A&A...587A..56M} were not able to constrain the surface gravity in their study. However, we note that surface gravity is not strongly constrained by our atmospheric model fitting. Our inferred radii from the fits to the first night and both nights spectra are consistent with the 1.55$\pm$0.1~R\textsubscript{Jup} radius inferred by \citet{2017ApJ...834..162G}, but our radius for the second night spectrum is slightly larger. Finally, all three of our inferred luminosities log(L/L$_\odot$) are lower than the $-2.76$$\pm$0.05~dex value measured by  \citet{2017ApJ...834..162G}. However, our luminosity from the first night is consistent within 1$\sigma$ with that found by \citet{2016A&A...587A..56M}, log(L/L$_\odot$)~=~$-2.81$$\pm$0.10 dex.

If we consider these results in full, the spectrum of HD~1160~B on our second night of observations does not appear to be consistent with the literature. If the differences between the spectra from each night at bluer wavelengths are due to astrophysical variability in the atmosphere of HD~1160~B, this highlights the impact that this can have on the results of fitting models to the atmospheres of substellar companions. Difficulties in fitting the spectrum of HD~1160~B have also been noted previously. When analysing the SPHERE spectra in the Y, J, and H bands, \citet{2020MNRAS.495.4279M} found that HD~1160~B has a spectrum that is not well matched by any spectra in current spectral libraries, and were only able to obtain good fits by considering the Y+J and H bands separately. Several studies of other substellar companions have also reported such issues when trying to fit their spectra, sometimes finding wide-ranging results depending on the wavebands considered \citep[e.g.][]{2020AJ....160..262S, 2021MNRAS.506.3224S, 2021AJ....161....5W, 2023MNRAS.525.1375W, 2024ApJ...961..210P}.

Simultaneous observations over a broad wavelength range with facilities such as JWST may help to explain these discrepancies between wavebands and further identify whether the differences in the spectrum of HD~1160~B between epochs are due to time variability \citep{2022PASP..134i5003H, 2022SPIE12180E..3NK, 2023ApJ...951L..20C, 2023ApJ...946L...6M, 2023PASP..135d8001R, 2024AJ....167..168M, 2024ApJ...966L..11P}. Ground-based high-resolution spectroscopy may further help us to determine its nature by resolving specific molecular lines that constrain effective temperature, surface gravity, and other physical properties \citep[e.g.][]{2013MNRAS.436L..35B, 2018arXiv180604617B, 2018A&A...617A.144H, 2019AJ....157..114B, 2023MNRAS.522.2145V}. High-resolution spectroscopy could also measure the $\varv \sin i$ of HD~1160~B, which would provide an independent and complementary upper limit of its rotation period \citep[e.g.][]{2014Natur.509...63S, 2016A&A...593A..74S, 2018NatAs...2..138B, 2020ApJ...905...37B, 2023A&A...670A..90P, 2021AJ....162..148W, 2022ApJ...937...54X, 2024A&A...682A..48L, 2024MNRAS.531.2356P}.

\subsubsection{The mass of HD~1160~B}\label{p3_ch4_companion_mass}
It is further possible to infer estimates for the mass of HD~1160~B using our luminosity estimates from Section \ref{p3_ch4_spectral_fitting} and values for the age of the HD~1160 system. We used the BT-Settl \citep{2014IAUS..299..271A,2015A&A...577A..42B} isochrones for this purpose, which are valid for brown dwarfs and low mass stars. To obtain mass estimates, we first interpolated over the model grid of each isochrone and then evaluated them at our luminosity values. As the resulting mass estimates are highly age-dependent and the age of the HD~1160 system is not well constrained, we carefully considered the range of age estimates in the literature and chose to use a 10-125~Myr range. This is a combination of the 20-125~Myr age range found by \citet{2017ApJ...834..162G} considering the properties of HD~1160~A and a range of evolutionary models, and the lower 10-20~Myr ages favoured by the results of \citet{2020MNRAS.495.4279M} (the former study also produced a range based on HD~1160~ABC together, but this was narrower 80-125~Myr). Our chosen range also covers the $\sim$120~Myr age that the HD~1160 system would be expected to have if it is a member of the Psc-Eri stellar stream, as suggested by \citet{2019AJ....158...77C}. \citet{2016A&A...587A..56M} did allow ages up to 300~Myr in their study, but ages this old appear to be ruled out by \citet{2017ApJ...834..162G}.

The resulting estimated mass ranges are shown in Table~\ref{table:p3_ch4_model_physical_parameters}. We also include the corresponding values for the mass ratio relative to HD~1160~A, $q$, assuming a stellar mass of 2.05~M$_{\odot}$ \citep[A1\,V,][]{2013ApJS..208....9P}. Older ages return higher mass values, and vice versa. Our full range of mass estimates covers 16-81~M\textsubscript{Jup}. This places HD~1160~B comfortably above the deuterium burning limit \citep[$\sim$11-16.3~M\textsubscript{Jup},][]{2011ApJ...727...57S}, but does not rule out the possibility that it is a low mass star above the hydrogen burning limit \citep[78.5~M\textsubscript{Jup},][]{2023A&A...671A.119C}. This is fully consistent with mass estimates in the literature, as we might expect given the broad age range assumed. \citet{2012ApJ...750...53N} estimated the mass of HD~1160~B to be 33{\raisebox{0.5ex}{\tiny$\substack{+12 \\ -9}$}}~M\textsubscript{Jup} upon its discovery, and \citet{2016A&A...587A..56M} found a mass range of 39-166M\textsubscript{Jup} based on their wider range of allowable system ages. \citet{2017ApJ...834..162G} later found a mass range of 35-90~M\textsubscript{Jup}. Finally, \citet{2020MNRAS.495.4279M} estimated a mass of $\sim$20~M\textsubscript{Jup} for HD~1160~B, which falls at the lower end of our range.

Our ability to precisely estimate the mass of HD~1160~B is severely limited by the highly uncertain age of the HD~1160 system, thus it will be difficult to further constrain the mass of this companion without either tighter constraints on its age or a dynamical mass measurement \citep[e.g.][]{2010ApJ...711.1087K, 2012ApJ...751...97C, 2016ApJ...827...23D, 2017ApJS..231...15D, 2019AJ....158..140B, 2022A&A...658A.145B, 2022A&A...668A.140R}.

\subsection{PEPSI characterization of HD~1160~A}\label{p3_ch4_discussion_pepsi}
In Section~\ref{p3_ch4_PEPSI_method_results}, we fitted BT-NextGen models to the PEPSI data of HD~1160~A and estimated its physical properties, finding T\textsubscript{eff}~=~9200$^{\,+\,200}_{\,-\,100}$\,K, $\varv \sin i$~=~96$^{\,+\,6}_{\,-\,4}$\,km~s$^{-1}$ and log(g)~=~3.5$^{\,+\,0.5}_{\,-\,0.3}$~dex. If we compare these physical properties to those of the best-fitting model found by using VOSA to fit the literature SED of HD~1160~A for the flux calibration process in Section~\ref{p3_ch4_spectral_extraction_fluxcalib}, the temperature and metallicity have the same values but the surface gravity here is lower than that of the log(g)~=~4.5~dex model found with VOSA. This difference in surface gravity does not significantly impact the flux calibration of the spectra of HD~1160~B in Section~\ref{p3_ch4_spectral_extraction_fluxcalib}, as the best-fitting model found with VOSA is convolved to the R$\sim$40 resolution of ALES prior to being used for this purpose. At this resolution, the two stellar models are indistinguishable and do not lead to differences in the HD~1160~B spectral fitting results.

If we assume the BT-Settl model grid spacing for log(g), 0.5~dex, as the uncertainty on the VOSA result, the two values are consistent within 1$\sigma$. We note that surface gravity is difficult to constrain with atmospheric models, and similarly good fits to the PEPSI data could also be obtained with slightly higher surface gravities. However, if HD~1160~A does have log(g)~=~3.5, this may be an indicator of youth, as older objects are likely to have higher surface gravities \citep[e.g.][]{2015A&A...577A..42B}.

Our measurement of $\varv \sin i$~=~96$^{\,+\,6}_{\,-\,4}$\,km~s$^{-1}$ is fully consistent with those previously found by \citet{2023AJ....165..164B}, who measured the $\varv \sin i$ of HD~1160~A at 3 different epochs using the Tull Coud\'e Spectrograph on the Harlan J. Smith telescope, finding values of $\varv \sin i$~=~96$\pm$10\,km~s$^{-1}$, 97$\pm$7\,km~s$^{-1}$, and 95$\pm$7\,km~s$^{-1}$ at each epoch, respectively.

Our derived values for the physical properties of HD~1160~A from the PEPSI spectrum correspond to an A1\,IV-V spectral type. This is a slightly later spectral type than the A0\,V classification found by \cite{1999mctd.book.....H} using photographic plates. \citet{2012ApJ...750...53N} previously noted that HD~1160~A is underluminous for its position on the HR diagram, based on it being an A0\,V star, and interpreted this as a sign of youth \citep[e.g.][]{1998ApJ...505..897J}. However, if HD~1160~A is an A1\,V star, this may partially account for this apparent underluminosity. As we also measured the $\varv \sin i$ of HD~1160~A, we further considered the alternative possibility that gravity darkening could help to explain this. If a star rotates rapidly it becomes oblate, leading to a greater radius and hence lower temperature and brightness at its equator compared to its poles \citep{2007Sci...317..342M, 2011A&A...533A..43E, 2022MNRAS.517.2942L}. The orbital inclination of HD~1160~B is almost edge-on \citep[92$^{+8.7}_{-9.3}$\textdegree,][]{2020AJ....159...63B}. Assuming that the stellar rotation axis of HD~1160~A is aligned with the orbit of HD~1160~B, rapid rotation would therefore lead to an apparent decrease in its luminosity as viewed from Earth. However, a typical A-type star has a much faster rotation (e.g. $\sim$190\,km~s$^{-1}$ for an A0 star) than our rotational velocity measurement $\varv \sin i$~=~96$^{\,+\,6}_{\,-\,4}$\,km~s$^{-1}$ \citep{1965Obs....85..166M, 2013A&A...557L..10N}. Thus, gravity darkening caused by rapid rotation cannot be the underlying cause of the underluminosity, if HD~1160~A is viewed approximately edge-on. If HD~1160~A is instead viewed pole-on (i.e. $i\sim0$), its true rotational velocity $\varv$ would be far faster and therefore lead it to appear brighter at its poles than its equator and hence to a relative increase in its luminosity as viewed from Earth. This indicates that gravitational darkening cannot explain any underluminosity of HD~1160~A, and that this is better accounted for by it being of a later spectral type than previously thought.

%% file: sections/09_conclusions.tex
\section{Conclusions}\label{p3_ch4_conclusions}
We present here a new study of the HD~1160 system using two nights of observations obtained with the Large Binocular Telescope. This work is divided into three parts: variability monitoring of red companion HD~1160~B with the dgvAPP360 coronagraph and the ALES IFS; a R$\sim$40 spectral characterization of HD~1160~B using the same data; and lastly a spectral characterization of host star HD~1160~A using R~=~50,000 high resolution spectroscopy obtained with the PEPSI spectrograph.

The variability analysis of HD~1160~B was conducted following the technique of gvAPP-enabled differential spectrophotometric monitoring recently presented by \citet{2023MNRAS.520.4235S}, who demonstrated this approach with the first night of observations used here. We first processed the LBT/ALES+dgvAPP360 data and extracted aperture photometry of both HD~1160~A and B, before combining the data in the wavelength dimension and dividing the companion flux by that of the star to produce a differential white-light curve for HD~1160~B spanning both nights. We then further detrended the light curve using a multiple linear regression approach. We find that we recover the high-amplitude $\sim$3.2~h periodic variability identified by \citet{2023MNRAS.520.4235S} in the first night, but that the second night light curve does not contain significant periodic variability, potentially indicating rapid time evolution in the atmosphere of HD~1160~B and highlighting the complexity of interpreting the light curves of high-contrast substellar companions. We also analysed the precision achieved in the detrended differential white-light curve on each night and found that the noise properties were similar. This suggests that gvAPP-enabled differential spectrophotometric monitoring achieves a repeatable precision at the few percent level over multiple epochs and that we do not reach the photon noise limit. Thus, a greater precision could be achieved in future studies if residual systematics in the differential light curves can be further mitigated using more advanced detrending approaches such as those using Gaussian processes \citep[e.g.][]{2012MNRAS.419.2683G, 2022MNRAS.510.3236P}.

We conducted our spectral characterization of HD~1160~B by instead combining the LBT/ALES+dgvAPP360 observations over the time sequence for each night, thereby producing 2.9-4.1~$\upmu$m spectra of the companion. These spectra are the first for this target in the mid-infrared and are therefore highly complementary to previous studies in the literature. We find that the spectrum of HD~1160~B from the second night is systematically fainter in the 3.0-3.2~$\upmu$m wavelength range than on the first night, which could be due to the intrinsic variability of the companion if this difference is astrophysical. We then fit these spectra with BT-Settl atmospheric models, considering each night separately and both nights together, and found that the results differ considerably depending on the data being fitted. Our effective temperature T\textsubscript{eff} estimates range from 2279{\raisebox{0.5ex}{\tiny$\substack{+79 \\ -157}$}}~K for the second night spectrum to 2794{\raisebox{0.5ex}{\tiny$\substack{+115 \\ -133}$}}~K on the first night. This first night T\textsubscript{eff} is consistent with the literature, but those derived from the second night and both nights combined spectra are cooler. Our inferred luminosities are lower than those in the literature, but our radius estimates are mostly consistent. Overall, we conclude that the spectrum of HD~1160~B on the second night of our observations is not consistent with the literature. The differences in the results obtained for each spectrum highlights the impact that variability can have on atmospheric model fitting for substellar companions. Simultaneous observations over a broad wavelength range with facilities such as JWST may help to resolve the ambiguities arising from these model fits and determine whether the differences in the spectrum of HD~1160~B between epochs are due to time variability.

By evaluating our luminosity estimates with BT-Settl isochrones over an age range of 10-125~Myr, we also estimated the mass of HD~1160~B. We report a 16-81~M\textsubscript{Jup} mass range, consistent with previous estimates in the literature. This places HD~1160~B comfortably above the deuterium burning limit, but also allows the possibility that it is a low mass star above the hydrogen burning limit.

Lastly, we performed a new characterization of host star HD~1160~A by comparing the R$\sim$50,000 high resolution spectrum obtained with PEPSI to BT-NextGen atmospheric models. We found values for the physical properties of HD~1160~A; T\textsubscript{eff}~=~9200$^{\,+\,200}_{\,-\,100}$\,K, log(g)~=~3.5$^{\,+\,0.5}_{\,-\,0.3}$, and $\varv \sin i$~=~96$^{\,+\,6}_{\,-\,4}$\,km~s$^{-1}$. This model corresponds to a spectral type of A1\,IV-V, which is slightly later than the literature A0\,V classification found by \cite{1999mctd.book.....H} using photographic plates. This may explain the apparent underluminosity of HD~1160~A previously noted by \citet{2012ApJ...750...53N}. By considering our rotational velocity $\varv \sin i$ measurement alongside the known near edge-on inclination angle of the HD~1160 system, we find that HD~1160~A rotates slower than the typical A-type star, and hence rule out gravitational darkening as the cause of any underluminosity.

Tighter limits on the age of the HD~1160 system or dynamical mass measurements of each component are key if the physical properties of HD~1160~A and B are to be constrained further. Observations over a broad wavelength range or at a high spectral resolution will also help to resolve the ambiguities in the spectrum of HD~1160~B, while additional epochs of ground-based differential spectrophotometric monitoring or high-precision space-based monitoring will shed light on its variability.

%% file: sections/10_acknowledgements.tex
\section*{Acknowledgements}\label{p3_ch4_ack}
The authors would like to thank the exoZoo team and Aneesh Naik for valuable discussions that improved this work. We also thank our anonymous referee whose comments helped us to improve this work. BJS and BB acknowledge funding by the UK Science and Technology Facilities Council (STFC) grant no. ST/V000594/1. BJS was also supported by the Netherlands Research School for Astronomy (NOVA). JLB acknowledges funding from the European Research Council (ERC) under the European Union's Horizon 2020 research and innovation program under grant agreement No 805445. This paper is based on work funded by the United States National Science Foundation (NSF) grants 1608834, 1614320, and 1614492. The research of DD and FS leading to these results has received funding from the European Research Council under ERC Starting Grant agreement 678194 (FALCONER).

We acknowledge the use of the Large Binocular Telescope Interferometer (LBTI) and the support from the LBTI team, specifically from Emily Mailhot, Jared Carlson, Jennifer Power, Phil Hinz, Michael Skrutskie, and Travis Barman. The LBT is an international collaboration among institutions in the United States, Italy and Germany. LBT Corporation partners are: The University of Arizona on behalf of the Arizona Board of Regents; Istituto Nazionale di Astrofisica, Italy; LBT Beteiligungsgesellschaft, Germany, representing the Max-Planck Society, The Leibniz Institute for Astrophysics Potsdam, and Heidelberg University; The Ohio State University, representing OSU, University of Notre Dame, University of Minnesota and University of Virginia. We gratefully acknowledge the use of Native land for our observations. LBT observations were conducted on the stolen land of the Ndee/Nn\=e\=e, Chiricahua, Mescalero, and San Carlos Apache tribes.

This publication makes use of VOSA, developed under the Spanish Virtual Observatory (\url{https://svo.cab.inta-csic.es}) project funded by MCIN/AEI/10.13039/501100011033/ through grant PID2020-112949GB-I00. VOSA has been partially updated by using funding from the European Union's Horizon 2020 Research and Innovation Programme, under Grant Agreement \textnumero 776403 (EXOPLANETS-A). This work has made use of data from the European Space Agency (ESA) mission {\it Gaia} (\url{https://www.cosmos.esa.int/gaia}), processed by the {\it Gaia} Data Processing and Analysis Consortium (DPAC, \url{https://www.cosmos.esa.int/web/gaia/dpac/consortium}). Funding for the DPAC has been provided by national institutions, in particular the institutions participating in the {\it Gaia} Multilateral Agreement. This publication makes use of data products from the Wide-field Infrared Survey Explorer, which is a joint project of the University of California, Los Angeles, and the Jet Propulsion Laboratory/California Institute of Technology, funded by the National Aeronautics and Space Administration. This publication makes use of data products from the Two Micron All Sky Survey, which is a joint project of the University of Massachusetts and the Infrared Processing and Analysis Center/California Institute of Technology, funded by the National Aeronautics and Space Administration and the National Science Foundation. This research has made use of the SIMBAD database and the VizieR catalogue access tool, operated at CDS, Strasbourg, France \citep{2000A&AS..143....9W, 2000A&AS..143...23O}. This research has made use of the NASA Exoplanet Archive, which is operated by the California Institute of Technology, under contract with the National Aeronautics and Space Administration under the Exoplanet Exploration Program. This research has made use of NASA's Astrophysics Data System. This research made use of SAOImageDS9, a tool for data visualization supported by the Chandra X-ray Science Center (CXC) and the High Energy Astrophysics Science Archive Center (HEASARC) with support from the JWST Mission office at the Space Telescope Science Institute for 3D visualization \citep{2003ASPC..295..489J}. This work made use of the whereistheplanet\footnote{\url{http://whereistheplanet.com/}} prediction tool \citep{2021ascl.soft01003W}. This work makes use of the Python programming language\footnote{Python Software Foundation; \url{https://www.python.org/}}, in particular packages including NumPy \citep{harris2020array}, SciPy \citep{2020SciPy-NMeth}, Astropy \citep{2013A&A...558A..33A, 2018AJ....156..123A, 2022ApJ...935..167A}, HCIPy \citep{2018SPIE10703E..42P}, PyAstronomy \citep{pya}, PynPoint \citep{2012MNRAS.427..948A, 2019A&A...621A..59S}, Photutils \citep{larry_bradley_2022_6385735}, scikit-learn \citep{scikit-learn}, scikit-image \citep{scikit-image}, statsmodels \citep{seabold2010statsmodels}, pandas \citep{mckinney-proc-scipy-2010, jeff_reback_2022_6408044}, and Matplotlib \citep{Hunter:2007}.

%% file: sections/11_data_availability.tex
\clearpage
\section*{Data Availability}\label{p3_ch4_ava}
The data from the LBT/ALES+dgvAPP360 and LBT/PEPSI observations underlying this article will be available in the Research Data Management Zenodo repository of the Anton Pannekoek Institute for Astronomy shortly after publication, at \url{https://doi.org/10.5281/zenodo.7051242}.